\documentclass[fleqn,usenatbib]{mnras}
\pdfoutput=1

\usepackage{newtxtext,newtxmath}
\usepackage[T1]{fontenc}

\DeclareRobustCommand{\VAN}[3]{#2}
\let\VANthebibliography\thebibliography
\def\thebibliography{\DeclareRobustCommand{\VAN}[3]{##3}\VANthebibliography}

\usepackage{longtable}
\usepackage{float}
\usepackage{graphicx}
\usepackage{amsmath}
\usepackage{caption}
\usepackage{subcaption}

\newcommand{\msun}{M$_\odot$}
\defcitealias{Schiappacasse2021}{SU21}
\defcitealias{Carretta2009u}{C09}
\newcommand{\teff}{T$_{\mathrm{eff}}$}

\title[n-capture/Li in NGC~6752]{Neutron-capture elements in NGC~6752 multiple populations}

\author[J. Schiappacasse-Ulloa et al.]{
J. Schiappacasse-Ulloa,$^{1,2}$\thanks{E-mail: joseluis.schiappacasseulloa@studenti.unipd.it}
S. Lucatello$^{2,3}$
\\
$^{1}$Dipartimento di Fisica e Astronomia, Universita’ di Padova, Vicolo dell' Osservatorio 3, I-35122, Padova, Italy\\
$^{2}$INAF–Osservatorio Astronomico di Padova, Vicolo dell’Osservatorio 5, 35122 Padova, Italy\\
$^{3}$Institute for Advanced Studies, Technische Universit{\"a}t M{\"u}nchen, Lichtenbergstra{\ss}e 2 a, 85748 Garching bei M{\"u}nchen\\
}

\date{Accepted XXX. Received YYY; in original form ZZZ}

\pubyear{2022}

\begin{document}
\label{firstpage}
\pagerange{\pageref{firstpage}--\pageref{lastpage}}
\maketitle

\begin{abstract}
Globular clusters have been widely studied in terms of light element variations present in their different stellar populations. However, the nature of the polluter(s) responsible for this phenomenon is still debated. The study of heavy elements and their relation to light ones can provide further constraints. In particular, we aim to explore the possible contribution of asymptotic giant branch stars of different stellar masses to the internal pollution in the cluster. We derive abundances of elements from different nucleosynthetic chains, such as Na, Mg, Ca, Sc, Cu, Y, and Ba. We did not find clear relations between the light s-process elements (represented by Y II) or heavy ones (represented by Ba II) with light elements (Li, Na or Al). This indicates that the polluter(s) responsible for the Na (Al) or Li production does not produce large amounts of Y II and Ba II. Furthermore, the comparison with models discards a possible significant contribution to the cluster pollution from AGB stars with masses lower than 5\msun. In addition, we found a potential CH-star in our sample.
\end{abstract}


%%%%%%%%%%%%%%%%%%%%%%%%%%%%%%%%%%%%%%%%%%%%%%%%%%

%%%%%%%%%%%%%%%%% BODY OF PAPER %%%%%%%%%%%%%%%%%%

\section{Introduction}

Globular clusters (GCs) are fossils from the early age of the Milky Way (MW). They have contributed to the formation of the Halo \citep{Martell2011} and likely the Bulge \citep{Lee2019}. Understanding the processes that lead to their formation, evolution and disruption/dissolution is an important ingredient in understanding the formation of these Galactic components. With the exception of some early work exploring the possibility that GCs were more complex objects \citep[e.g.,][]{Norris1981}, up to the early 21st century, they were thought to be a classic example of a simple stellar population, i.e. a coeval group of stars, with the same initial chemical composition. This view has changed in the last couple of decades. Analysis of colour-magnitude diagrams (CMDs) of GCs have shown parallel sequences along every evolutionary phase \citep[e.g.][]{Milone2017} and high-resolution spectroscopy revealed different chemical patterns all the way down to the main sequence, a peculiar characteristic of these objects. A milestone was given by \citep{Carretta2009g} and \citep[][hereafter C09]{Carretta2009u} who for the first time found, in a large sample of GCs, the star-to-star variations in light elements showing that this phenomenon is a generalised phenomenon in Galactic CGs, to the point of having been suggested as their defining feature.

Nowadays, there is general agreement that these photometric and spectroscopic pieces of evidence reflect the presence of multiple stellar populations (MSP). In the most accepted scenario, there is a first generation (FG) of stars formed from pristine material. When a fraction of these stars evolved, ejecting gas rich in elements that were processed in their interiors, which is mixed with different amounts of pristine material, forming a second-generation (SG) of stars with altered composition \citep[][]{Bastian2018,Gratton2019}. These stellar generations can be distinguished by their chemical signatures: standard Population II composition for the FG, depleted (enhanced) C, O, and Mg (N, Na, and Al) for stars belonging to the SG. These particular patterns form the anti-correlations (e.g., C-N, Na-O and Mg-Al) found in GCs. While the MSP phenomenon has been widely studied, the origin of the polluters responsible for the light-element variations in SG stars is not clear yet. Different candidates have been proposed as responsible for chemical pollution, although none of them has been able to reproduce the full range of observables. The most popular candidate polluters are fast-rotating massive stars \citep[][]{Decressin2007}, massive binaries \citep[][]{deMink2009}, and intermediate-mass ($\sim$4-8~\msun) asymptotic giant branch \citep[AGB;][]{Ventura2001} stars.

To clarify the nature of the polluter, it is key to satisfactorily link their nucleosynthesis output and the abundance patterns found in GCs. The light-element variations have been found in all Galactic GCs examined in detail, and they are attributed to the nucleosynthesis during the hot H-burning \citep[][]{Denisenkov1989}: C-N, Na-O and Mg-Al anti-correlated with each other. On the other hand, only a subset of GCs shows clear evidence of Fe-peak elements variations, such as $\omega$-Cen \citep{Villanova2014}, M54 \citep{Bellazzini_2008}, and NGC~1851 \citep{Carretta_2011} among others. K, Ca, and Sc, have also been shown to exhibit variations in some clusters \citep[e.g., NGC~2419;][]{Cohen2012}. \citet[][]{Ventura2012} claimed that AGB stars could produce elements like K or Ca if they burn H at high enough temperatures. These elements can help us to constrain the nature of the cluster polluters, and several recent studies have highlighted their abundances. Recently \citet[][]{Carretta2021} analysed Ca, and Sc in a large set of GCs finding an enrichment of Ca in NGC 4833, NGC 6715, NGC 6402, NGC 5296, NGC 5824, and $\omega$-Cen compared to field stars of similar metallicities. In those GCs, they found higher Ca abundances in stars belonging to SG stars.

Another element which has been the object of interest in this context is Li. A fragile element, it is easily burned into $^{4}$He at low temperatures ($\sim2.5\times10^{6}$~K). In hot H-burning processed material, which must have reached considerably higher temperatures to activate the NeNa and MgAl cycles typical of the observed SG abundances, lithium should be obliterated. Therefore, stars belonging to the SG should contain low amounts of Li, strongly anti-correlating with the Na and Al abundances. Several studies have looked at the Li content along with other light elements involved in the MSP phenomenon. Surprisingly, some GCs such as, NGC~6121 \citep{Dorazi2010,Mucciarelli20111}, NGC~6218 \citep{dorazi2014}, and NGC~362 \citep{Dorazi2015} showed considerable Li abundance among SG stars. In this context, NGC~6752 is an interesting GC which shows clear Na-O and MgAl anti-correlations \citep{Carretta2009u}. Moreover, it has a metallicity of -1.56\citep{Carretta2009c} with some iron spread among FG stars \citep{Legnardi2022}. Lately, in a recent paper \citep[][hereafter SU21]{Schiappacasse2021}, we analysed 217 stars from the turn-off (TO) to the bottom red giant branch (RGB) in the GC NGC~6752. We found FG stars with high Li and low Al, and SG stars with low Li and high Al, but in addition, we found a fraction of SG stars with high Al and Li. Since AGB stars, through the Cameron-Fowler mechanism \citep{Cameron1971}, are the only polluters capable to produce Li along with the other light-element variations, these results strongly argue in favour of their contributions to the chemical pollution associated with the MSP phenomenon. However, these results do not discard other sources of pollution in addition to the AGB.

Heavier, neutron-capture (n-capture) species have been the object of limited investigations so far: studies have shown that they display quite homogeneous abundances in most clusters \citep[e.g.,][]{dorazi2010_neutron,Cohen2011}, although, in some GCs there is evidence of considerable spread in the slow n-capture\footnote{It refers to elements produced by capturing a neutron at a slower rate than the $\beta^{-}$ decay reaction.} (s-process) elements e.g., NGC~7089 \citep{Yong2014}, and M~22 \citep{Marino2009}, and in rapid n-capture e.g., M15 \citep{Sobeck2011}. Some of the classes of polluters, specifically AGB stars, are expected to produce different amounts of s-process elements depending on their masses. Then, extending the study of the chemistry of GCs to heavy elements, allows us to better understand the nature of the polluters. For example, it has been shown that fast-rotating massive stars (M$\geq$8-12~\msun) can increase drastically the production of s-process elements, such as Sr, Ba, or La, compared to their non-rotating counterparts \citep[][]{Shingles2014}. Nevertheless, it would produce them at later stages of their evolution, releasing their material during the core-collapse supernova explosion, which cannot be held in most of the cluster \citep[][]{Dorazi2013}. On the other hand, low- and intermediate-mass AGB stars ($\sim$1.2-8.0~\msun) produce s-process elements during their thermal pulses. During the third dredge-up, the star can make different s-process elements from the lightest (ls; Sr, Y, Zr) to the heavier (hs; Ba, La, Ce) up to Pb. The AGB stars yields depend strongly on their initial stellar mass (and metallicity), with AGB stars with lower masses more efficiently producing heavier s-process elements (e.g., Ba, and La), and lighter s-process elements (e.g. Sr, and Y) by more massive AGB stars \citep[][]{Karakas2014}. Then, by analyzing the behaviour of s-process elements within the cluster populations one can tease out a further piece of the puzzle in the study of the nature of the polluters.

In the present paper, we extend the analysis done by \citetalias{Schiappacasse2021} to O, Mg, Ca, Sc, Cu, and n-capture elements aiming to fully characterise the different populations found in NGC~6752 and to constrain the nature of the polluter(s) responsible for its chemical peculiarities. The structure of the paper is the following: In \S\ref{Sec:Observation}, we describe the target selection and observation. In \S\ref{Sec:Data Analysis}, we described the determination of stellar parameters, and the analysis of the observational uncertainties. We show our results Cu, Y, Ba, and Eu and we discuss potential relation with Li in \S\ref{sec:results}. Finally, we briefly summarize our findings in \S\ref{Sec:Conclusion}.

\section{Data collection and reduction}
\label{Sec:Observation}

We collected the spectra of the GC NGC~6752 in different stellar evolution stages from the TO up to the RGB-bump. We analysed the 158 spectra with the suitable spectral range for our purposes. A large part of them (67 stars) were previously used by \citet{Gruyters2014} and later by \citetalias{Schiappacasse2021}. Those spectra were kindly provided by the authors. The remaining 91 stars are from the archival Gaia-ESO collection, and they were also used by \citetalias{Schiappacasse2021}. In summary, the spectra are a mix of: \\
11 FLAMES/UVES spectra (4768-5801~\r{A} and 5822-6830~\r{A} with $R \equiv$ 47000 each), \\
147 FLAMES/GIRAFFE (HR15N: 6444-6816~\r{A} with $R \equiv$~19200), \\
66 stars from HR10 (5330-5620~\r{A}, $R\equiv \lambda/\Delta\lambda$=~21500), \\
105 stars from HR11 (5600-5840~\r{A}, $R\equiv \lambda/\Delta\lambda$=~29500), \\
50 stars from HR13 (6120-6405~\r{A}, $R\equiv \lambda/\Delta\lambda$=~26400), \\
126 stars from HR15 (6600-6960~\r{A}, $R\equiv \lambda/\Delta\lambda$=~21300). 

The normalization of the continuum, the radial velocity measurement, and the spectra shift to the rest-frame were performed with \texttt{iraf}, and are described, along with the membership selection in detail in \citetalias{Schiappacasse2021}.

\section{Data Analysis}
\label{Sec:Data Analysis}

We adopt the atmospheric parameters provided by \citetalias{Schiappacasse2021}. 
Briefly, we used Str\"{o}mgren photometry \citet{Grundahl1999} to derive the photometric effective temperatures (T$_{\mathrm{eff}}$) using the colour-temperature relations reported by \citet{Korn2007}, who modified the temperature scales given by \citet{Alonso1996} and \citet{Alonso1999} for evolved and unevolved stars, respectively. We used the isochrone from \citet{Bressan2012} with 13.5~Gyr \citep{Gruyters2013}, and [Fe/H]=-1.56 dex \citep{Carretta2009c} to determined $\log g$ from photometric T$_{\mathrm{eff}}$. Finally, we got the microturbulence velocity (v$_{m}$) using the relation\footnote{v$_{t} = 2.22 - 0.322$ $\log g$} given by \citet{Gratton1999}. For a detailed description of the determination of the stellar parameters and the values we used, as well as the observational uncertainties, we refer the reader to the mentioned paper.

\subsection{Abundance Determinations}

We derived the Na, Mg, Ca, Sc, Cu, Y, and Ba abundances. Moreover, we derived upper limits for O and Eu abundances. While for UVES spectra, all these species were analysed, we measured them in GIRAFFE spectra whenever the order was available: HR10 (Mg: 5528\AA, and Y: 5509\AA), HR11 (Na: 5682\AA~-5688\AA, Mg: 5711\AA, and Cu: 5105 \AA), HR13 (O: 6300 \AA, and Ba: 6141\AA), HR15 (Ba: 6496\AA, and Eu: 6645\AA). We adopt as solar abundances those reported in \citet{Asplund2009}, noting that our solar analysis yields very similar results. Individual abundance determinations were done as follows. 

We used the model atmospheres grids from \citet{Kurucz1992} for our abundance determination. In particular, Mg, Ca, and Sc were measured automatically via equivalent widths (EW) using \texttt{ARES} \citep{Sousa2007}. The line list is the one reported in \citet{Dorazi2015}. On average, \texttt{ARES} could detect 17, 7 and 2 Ca I, Sc II and Mg I lines respectively. Lines with large fitting errors or that resulted in strongly discrepant abundances were checked manually using \texttt{iraf}. Abundances were derived from these EWs using \texttt{abfind} driver from \texttt{pyMOOGI}\footnote{Github site: \url{https://github.com/madamow/pymoogi}} \citep{Adamow_2017}.

The abundances derivation for O, Na, Cu, Y, Ba, and Eu were done through spectral synthesis using \texttt{pyMOOGI} with its driver \texttt{synth}, which is a 1D LTE\footnote{local thermodynamic equilibrium} line analysis code. The line lists for this method were generated with \texttt{linemake} code\footnote{Github site: \url{https://github.com/vmplacco/linemake}} \citep{Placco2021}, which consider hyperfine splitting for Ba II \citep{Gallagher1967}, Cu I\footnote{\url{http://kurucz.harvard.edu/atoms.html}} \citep{Kurucz1995}, and Eu II \citep{Lawler2001}. We assumed solar isotopic ratios from \citet[][]{Asplund_isoratios} for Cu, Y, Ba and Eu. Although the solar isotopic ratios for these elements are likely not appropriate for Population II stars, this will have a negligible effect on our results. In fact, in the present paper, we only use yellow/red Ba lines for which the isotopic split is quite small considering our resolution.

We analysed the O atomic line at 6300\AA and the Na doublet at 5682-5688\AA.  Both O and Na abundances were derived via spectral syntheses: because of the line weakness \texttt{ARES} was either unable to measure the EWs or returned very large associated errors. In fact, for O, no detection was possible even via synthesis, hence only upper limits were derived. We applied non-LTE corrections to the Na abundances using the corrections reported on INSPECT data base\footnote{\url{http://inspect-stars.com/}}, which is based on those provided by \citet{Lind2011}.

Ba II lines (5853\AA, 6141\AA, and 6496\AA), Y II (5087\AA, 5200\AA, and 5509\AA), Cu I (5105\AA) and Eu II (6645\AA) are lines affected by isotopic and hyperfine splitting, and the determination of the abundances was done by matching the synthetic spectra with the observed one. 

\subsection{Uncertainties associated with derived abundances}

The observational uncertainties are the contribution of the direct abundance determination and the uncertainty arising from the uncertainties associated with the atmospheric parameters adopted. The first is either by the error on the best-fit determination (for abundances derived via synthesis) or line-by-line abundance scatter (for abundances determined with EWs analysis). The second is derived by evaluating the variation of the abundances to the change in each of the parameters (T$_{\mathrm{eff}}$, $\log g$, v$_{m}$, and [Fe/H]). 

\begin{figure}
	\includegraphics[width=\columnwidth]{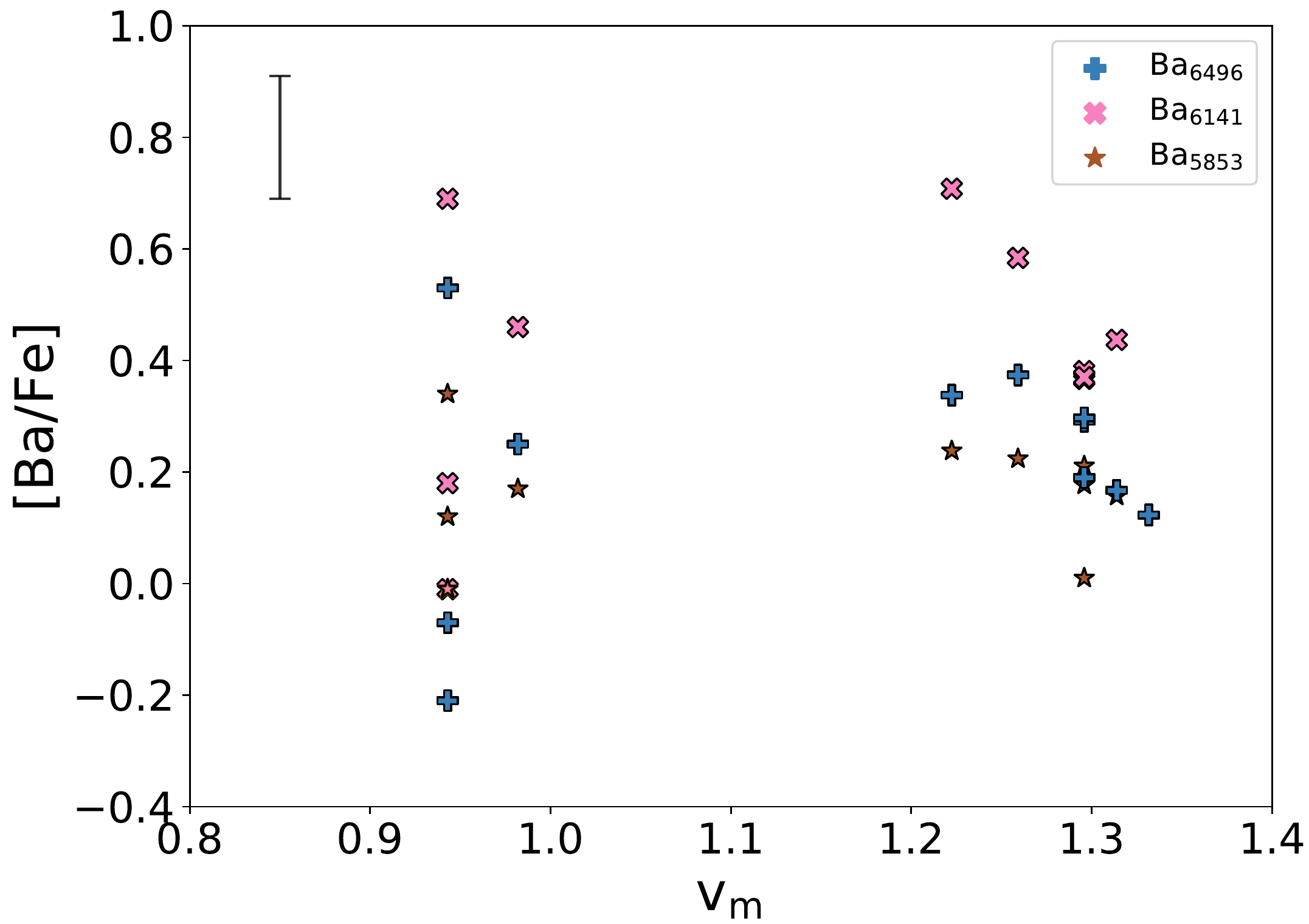}
    \caption{Ba abundances as a function of v$_{m}$ for our UVES sample. Three different symbols represent the Ba lines at 5853\AA, 6141\AA, and 6496\AA.}
    \label{fig:AllBaVm}
\end{figure}

In this context, it is worth mentioning that the Ba abundances, which are based on three rather strong lines, show considerable sensitivity to the adopted v$_{m}$. This is a common finding in cool giants, as discussed, e.g., by \citet[][]{Worley2013}. This effect can be clearly seen in the high sensitivity of Ba abundances to the variations of this parameter reported in Table \ref{tab:sensitivity_matrix}, and in Fig. \ref{fig:AllBaVm}, which shows the distribution of the abundance obtained for each line as a function of v$_{m}$. We explored averaging Ba abundances weighted by their respective errors using the different combinations of lines to minimise this effect and concluded that the best combination is indeed the use of all three. We opted to use from hereinafter all the tree lines for our final abundance, due to the reduction of both the spread and the lessening of the v$_{m}$ dependence. Similar considerations apply to the Y II lines used to derive [Y/Fe]II abundances. These dependencies are further discussed later in this paper. Cu abundances do not show any significant trend with v$_{m}$, which is expected for weak lines. 

To derive the sensitivity of the derived abundances to atmospheric parameters, we selected two stars as representative stars of the whole sample, being an unevolved (\#6003089) and an evolved (\#5958306) star. The perturbations in stellar parameters to estimate the sensitivity of the element are $\Delta$T$_{\mathrm{eff}} =  100$ K, $\Delta$$\log g=  0.2$ dex, $\Delta$v$_{m}$$= 0.1$ km s$^{-1}$, and $\Delta$[Fe/H]$ =  0.1$ dex. 
The abundance errors are based on the sensitivity matrix shown in Table \ref{tab:sensitivity_matrix}.

Table \ref{tab:abundance_table} summarises the abundances obtained for each element and associated uncertainty.

\begin{table*}
   \centering
   \caption{Sensitivity matrix for Na, Mg, Ca, Sc, Cu, Y, and Ba on representatives of TO/SGB star (\#19102677-6003089) and RGB star (\#19102025-5958306).} 
   \label{tab:sensitivity_matrix}
\begin{tabular}{cccccccccccc}
       \hline
       \hline
   &              &           & 19102677-6003089  &          &         &  &              &           & 19102025-5958306  &          &         \\
   \hline
   & Best$_{fit}$ & $\Delta$T$_{\mathrm{eff}}$ & $\Delta$$\log$ g    & $\Delta$[Fe/H] & $\Delta$v$_{m}$ &  & Best$_{fit}$ & $\Delta$T$_{\mathrm{eff}}$ & $\Delta$$\log$ g    & $\Delta$[Fe/H] & $\Delta$v$_{m}$ \\
   &              & (100 K)         & (0.2 dex)         & (0.1 dex)      & (0.1 km s$^{-1}$)    &  &              & (100 K)         & (0.2 dex)         & (0.1 dex)      & (0.1 km s$^{-1}$)       \\
       \hline
Na & 0.10         & 0.07      & 0.06    & 0.06     & 0.01    &  & 0.05         & 0.04      & 0.00    & 0.00     & 0.00    \\
Mg & 0.15         & 0.05      & 0.03    & 0.00     & 0.01    &  & 0.05         & 0.08      & 0.04    & 0.00     & 0.01    \\
Ca & 0.10         & 0.05      & 0.01    & 0.00     & 0.01    &  & 0.05         & 0.05      & 0.03    & 0.01     & 0.02    \\
Sc & 0.10         & 0.03      & 0.07    & 0.00     & 0.01    &  & 0.10         & 0.03      & 0.07    & 0.01     & 0.01    \\
Cu & 0.10         & 0.10      & 0.01    & 0.01     & 0.00    &  & 0.05         & 0.11      & 0.01    & 0.02     & 0.01    \\
Y  & 0.10         & 0.05      & 0.06    & 0.02     & 0.01    &  & 0.10         & 0.06      & 0.07    & 0.01     & 0.02    \\
Ba & 0.10         & 0.07      & 0.04    & 0.01     & 0.05    &  & 0.10         & 0.07      & 0.05    & 0.01     & 0.07    \\
       \hline
       \hline
\end{tabular}
\end{table*}

\setlength{\tabcolsep}{3.0pt}
\begin{table*}
   \centering
   \caption{Abundances and errors reported in this work for our sample. The full table will be available online.}
   \label{tab:abundance_table}
\begin{tabular}{ccccccccccccccccccccccccccc}
       \hline
       \hline
ID    & f & {[}O/Fe{]}  & f &{[}Na/Fe{]} & $\sigma$ & f & {[}Mg/Fe{]} & $\sigma$ & {[}Ca/Fe{]} & $\sigma$ & {[}Sc/Fe{]} & $\sigma$ & f &{[}Cu/Fe{]} & $\sigma$ & f & {[}Y/Fe{]} & $\sigma$  & f & {[}Ba/Fe{]} & $\sigma$ & f & {[}Eu/Fe{]}  \\
      &  & dex    &   & dex      &   &          & dex         &          & dex         &          & dex     &  &   &  dex         &       &   & dex        &           &  & dex         &       &   & dex         \\
        \hline
3860 &             & 9.99      & <             & 0.02        & 0.11 &                         & 0.06        & 0.27 & 9.99        & 9.99 & 9.99        & 9.99 &                         & 9.99       & 0.1  & <            & -0.2       & 0.11 & <             & 9.99       & 0.11 & <             & 9.99       \\
3849 &             & 9.99      & <             & 0.25        & 0.11 &                         & 0.31        & 0.11 & 9.99        & 9.99 & 9.99        & 9.99 &                         & 9.99       & 0.1  & <            & 0.3        & 0.11 & <             & 9.99       & 0.11 & <             & -0.1        \\
3790 &             & 9.99      & <             & -0.42       & 0.11 & <             & 0.35        & 0.11 & 9.99        & 9.99 & 9.99        & 9.99 &                         & 9.99       & 0.1  & <            & 9.99      & 0.11 & <             & 9.99       & 0.11 & <             & 0.1         \\
3778 &             & 9.99      & <             & -0.19       & 0.11 &                         & 0.2         & 0.11 & 9.99        & 9.99 & 9.99        & 9.99 &                         & 9.99       & 0.1  & <            & 0.1        & 0.11 &                         & 0.25        & 0.12 & <             & 9.99       \\
3747 &             & 9.99      & <             & -0.09       & 0.11 &                         & 0.17        & 0.11 & 9.99        & 9.99 & 9.99        & 9.99 &                         & 9.99       & 0.1  & <            & 0.48       & 0.11 & <             & 9.99       & 0.11 & <             & 9.99       \\
       \hline
       \hline
\end{tabular}
\end{table*}

\subsection{Fe spread}

\citet[][]{Milone2017} developed a two-colour diagram so-called chromosome map to disentangle the different populations in GCs. They have found that FG stars display an extended sequence in the diagram. \citet[][]{Marino2019} claimed it could be either given by a He- or Fe-spread among FG stars, meaning those stars were not as homogeneous as was thought. \citet[][]{Legnardi2022} analysed a large sample of GCs and they determined the extended FG sequence was produced due to an iron dispersion among their members. In particular, they estimated the Fe-dispersion in NGC~6752 to be 0.106$\pm$0.017.

\begin{figure*}
	\includegraphics[width=\textwidth]{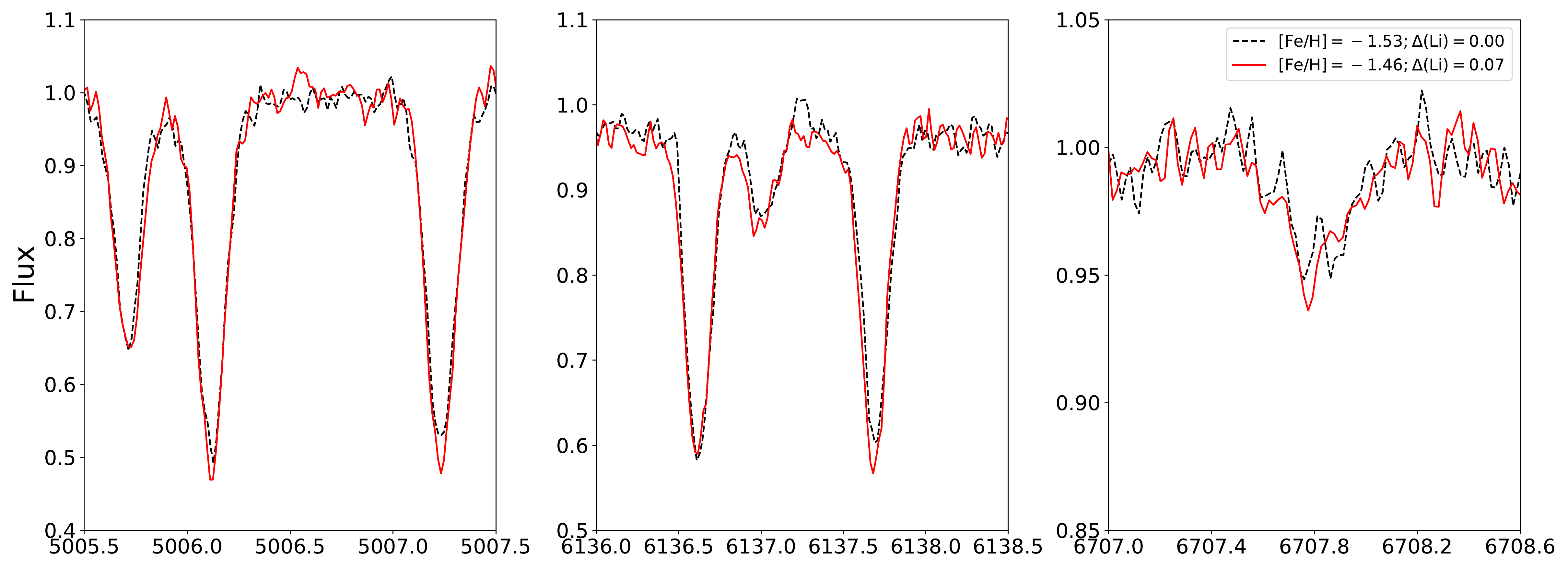}
    \caption{Comparison of Fe lines for two stars with similar stellar parameters. The red solid and the black dashed line represent two evolved stars with a T$_{\mathrm{eff}}$ of 5257 K and 5232 K, respectively.}
    \label{fig:comp_Fe}
\end{figure*}

We performed Fe abundance determination for our UVES sample, which has a broader wavelength range than the GIRAFFE ones to analyse the Fe-dispersion among FG stars. For those stars, we found an internal iron variation of 0.12$\pm$0.05 dex, which is in principle good agreement with the spread reported by \citet[][]{Legnardi2022}, but at the same time is also consistent with no Fe spread. To investigate this further, we compared the spectra of FG stars with similar stellar parameters with resulting different Fe. In our UVES sample, we have four FG stars with a difference in \teff~ of about 25 K.
Fig.\ref{fig:comp_Fe} shows a limited wavelength range with Fe lines for a graphical comparison of the two stars with the most different iron abundance. The [Fe/H] in our four stars range from -1.53$\pm$0.05 dex to -1.46$\pm$0.05 dex, meaning that considering the associated errors, we did not find a significant difference in their iron abundances. Nevertheless, due to our small sample, it must be taken with caution.

\section{Abundance Patterns}
\label{sec:results}

\subsection{Oxygen, Sodium, Magnesium, and Lithium}
\label{subsec:ONaMgLi}

O, Na, Mg, and Al have been used as tracers of the MSP phenomenon. GCs display the well-known anti-correlations among pairs of these elements, which are thought to reflect different nucleosynthetic processes inside of the cluster polluters. Hot H-burning results in the production of Na by proton-capture processes through the NeNa chain and of Al, through the AlMg chain, while O and Mg get depleted. However, Mg gets a more modest depletion than Al because the Mg number density is much higher than the one for Al. 

\begin{figure}
	\includegraphics[width=\columnwidth]{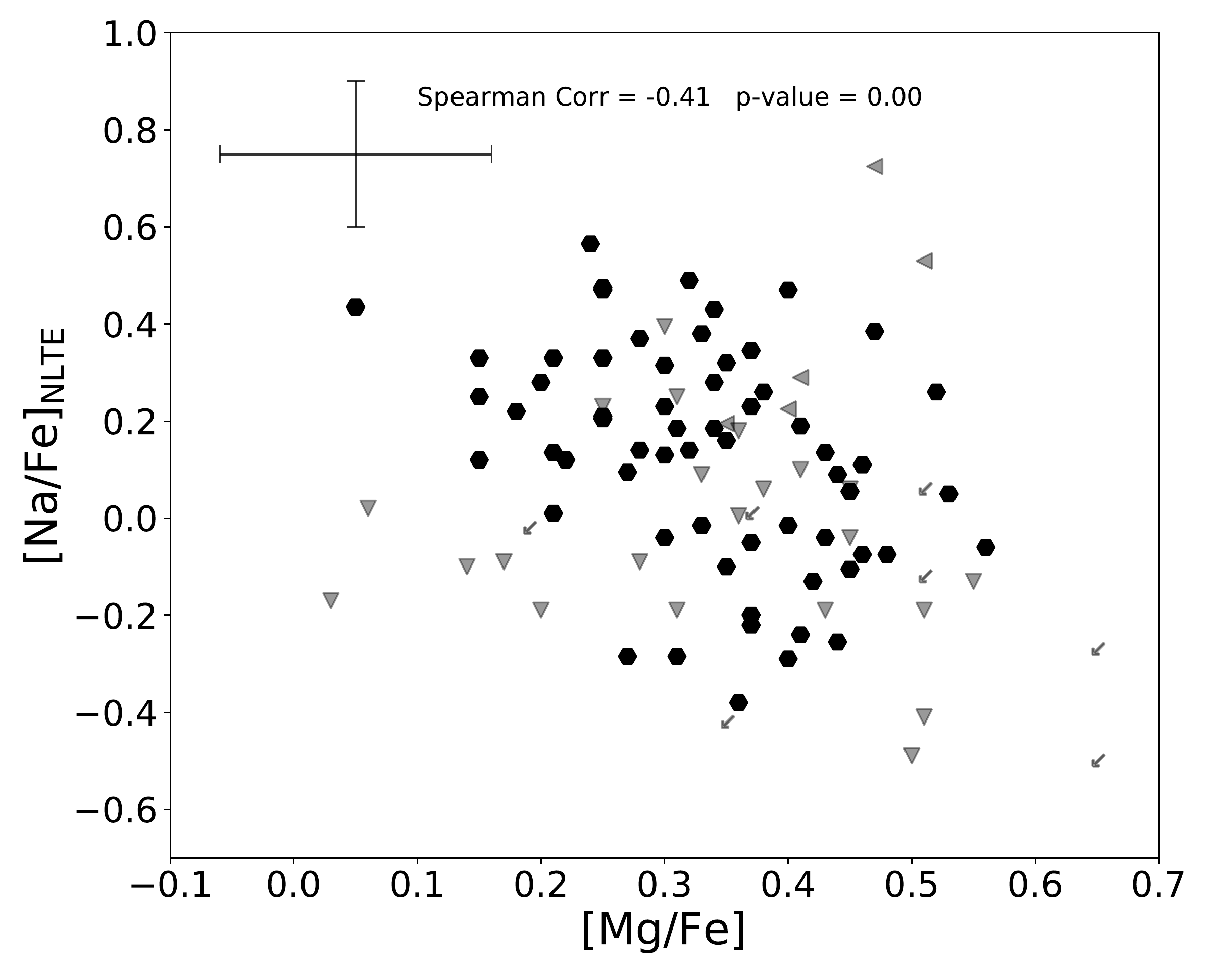}
    \caption{Mg-Na distribution in our sample. Symbols represent actual measurements (circles), Mg upper limits (left-pointing triangles), Na upper limits (down-pointing triangles), and the upper limits in both elements are represented by arrows.}
    \label{fig:MgNa}
\end{figure}

We analysed the spectroscopic indicators associated with the MSP phenomenon present in GCs. We note the Na-O anti-correlation is not significant in our sample, as our oxygen measurements are scarcely significant upper limits. Fig.\ref{fig:MgNa}, shows Na along with the Mg abundances. If both NeNa and AlMg chains were activated, SG stars are expected to be Na enhanced and Mg depleted with respect to their FG counterparts. In the figure, these two species are consistent with a statistically significant moderate negative correlation (Spearman corr. = -0.41 and p-value$\sim$0.0) and display a spread larger than the associated error. In particular, we found a [Na/Fe] range of at least 1.1 dex, which is in excellent agreement with the [Na/Fe] range ($\sim$1.1 dex) reported by \citet[][]{Carretta2007} for this cluster. On the other hand, the Mg spread is more modest ($\sim$0.50 dex), and it agrees with the Mg spread ($\sim$0.55 dex) reported by \citet[][]{Carretta2012}.

\begin{figure}
	\includegraphics[width=0.95\columnwidth]{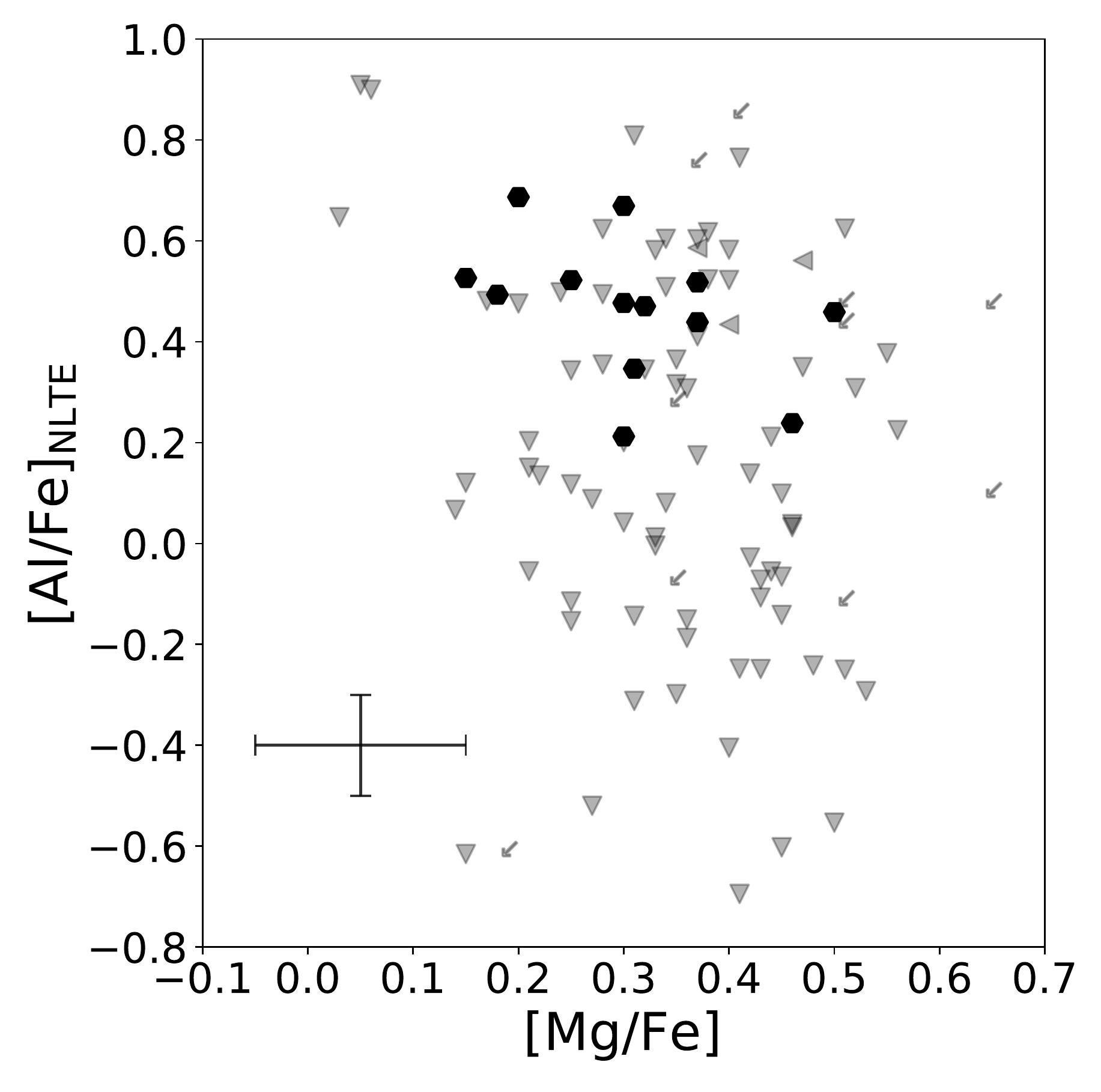}
    \caption{Mg-Al anti-correlation in our sample. Symbols follow the same description as Figure \ref{fig:MgNa}}
    \label{fig:MgAl}
\end{figure}

In addition, we used the Al abundances reported in \citetalias{Schiappacasse2021} and examined their relation to Mg. \citetalias{Schiappacasse2021} reported Al abundance measurements for 34  stars and upper limits for 183 more, for which we analysed the Mg abundances whenever possible. Our results are shown in Fig. \ref{fig:MgAl}. Although the results are highly dominated by upper limits, the actual measurement in both elements draws a statistically significant moderate negative correlation (Spearman corr.=0.59 and p-value=0.03), which is consistent with the anti-correlation reported in the literature. 

\begin{figure}
     \centering
     \begin{subfigure}[b]{\columnwidth}
         \centering
         \includegraphics[width=\columnwidth]{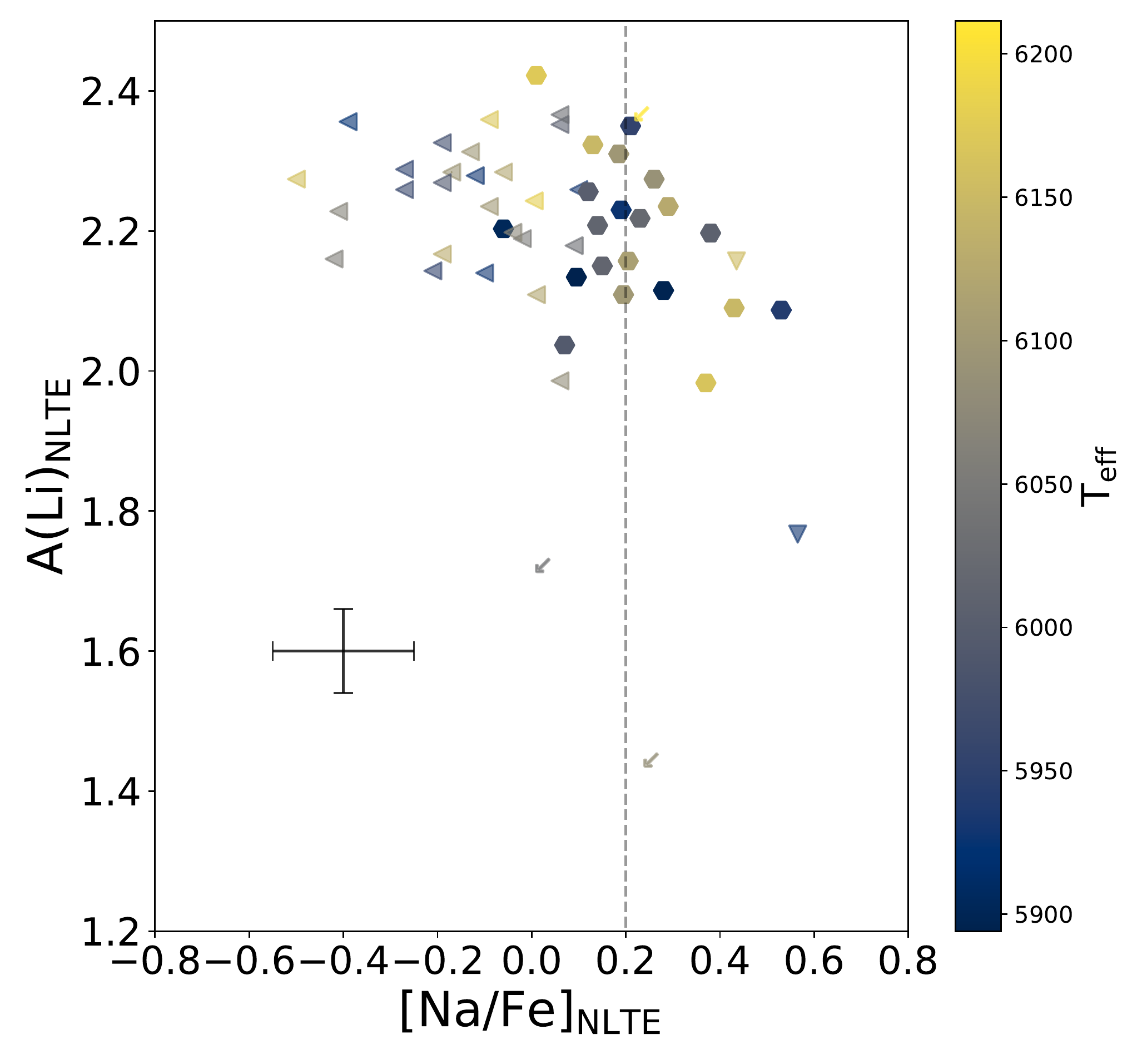}
     \end{subfigure}
     \vfill
     \begin{subfigure}[b]{\columnwidth}
         \centering
         \includegraphics[width=\columnwidth]{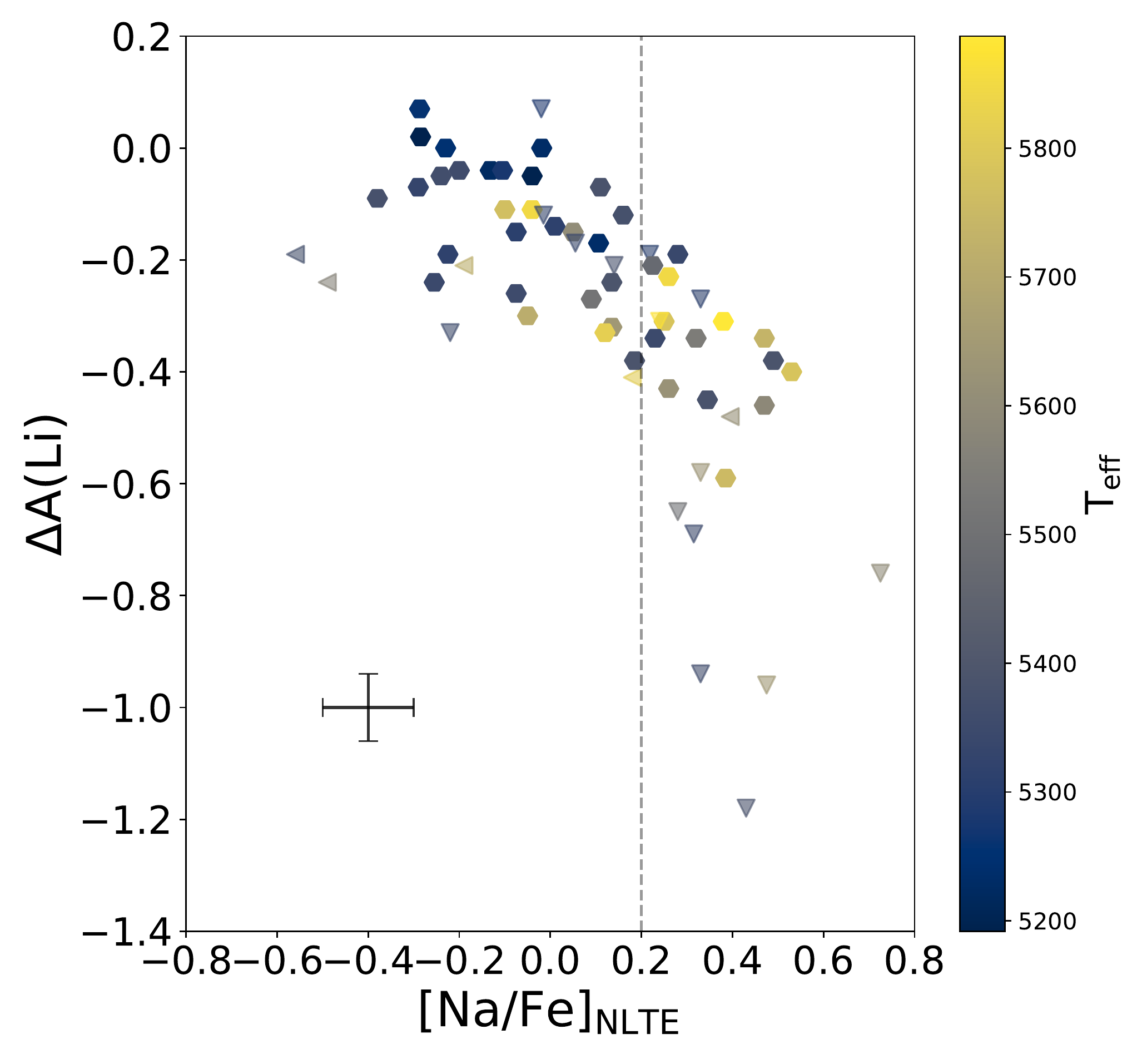}
     \end{subfigure}
        \caption{Li/$\Delta$A(Li) as a function of the Na abundance for unevolved (upper panel), and evolved stars (lower panel). The dashed line indicates the threshold [Na/Fe] value which splits the two generations in NGC~6752. Symbols follow the same description as the Figure \ref{fig:MgNa}, which are colour-coded depending on their \teff}
        \label{fig:LiNa_teff}
\end{figure}

\citetalias{Schiappacasse2021} reported abundances of Li in the same sample, where they analysed its behaviour from the TO to lower RGB. They define the term $\Delta$A(Li)\footnote{$\Delta$A(Li) is the difference between the expected Li abundance from stellar evolution models and the Li abundance measured by \citetalias{Schiappacasse2021}. See the referred paper for details.} to account for the Li variation due to the evolution of the stars, and single out the effect of MSP. Figure \ref{fig:LiNa_teff} shows the Li abundance as a function of Na for unevolved (upper panel), and evolved stars (lower panel), being defined as unevolved stars the ones with $\log g$>3.80. The dashed line indicates the threshold [Na/Fe] value reported by \citet[][]{Carretta2010} to split the two generations for NGC~6752. Symbols were colour-coded according to the stars' \teff. In both panels is clear the presence of an anti-correlation, where the stars get depleted along with the Na enrichment. Confirming the findings reported by \citetalias{Schiappacasse2021}, there are a few stars that belong to the SG that show a high Li abundance, which in some cases reaches the Li content of FG stars. The last indicates the contribution of Li pollution from AGB stars. As was mentioned by \citet[][]{Dantona2019}, the chemical pattern of SG star with 0.2 dex$\lesssim$[Na/Fe]$\lesssim$0.4 dex, could be explained by AGB stars of different masses. 

Unlike \citet{Mucciarelli20111}, who found that Li abundances of SG stars in the GC M4 were the same as in FG stars, we detected in both panels, that SG stars have on average a lower Li abundance than the FG ones. We did not detect any obvious dependence of this effect with stars' \teff. Finally, we note that there seems to be a larger dip in Li between FG and SG stars among evolved stars with respect to unevolved ones, which could be due to an under-correction of the evolutionary depletion.

\subsection{Calcium and Scandium}
Ca is an $\alpha$-element produced in SN-II \citep{Woosley_1995}, while Sc is often categorised as an iron-peak element, it is produced mostly in SN-II \citep{Battistini2015}. These elements have only recently started to be considered in the framework of MSP. \citet[][]{Carretta2021} analysed the Mg, Ca, and Sc in a large sample of GCs finding Ca excess with respect to the field stars in a handful of them (NGC 4833, NGC 6715, NGC 6402, NGC 5296, NGC 5824, and $\omega$-cen). They quantified that Ca excess, through a Kolmogorov–Smirnov test, getting statistically robust results. They claimed that such Ca excess could be produced either by a common kind of star in all the GCs being activated under specific conditions or by the presence (or absence) of an on-off mechanism in a peculiar kind of star. 

\begin{figure*}
	\includegraphics[width=\textwidth]{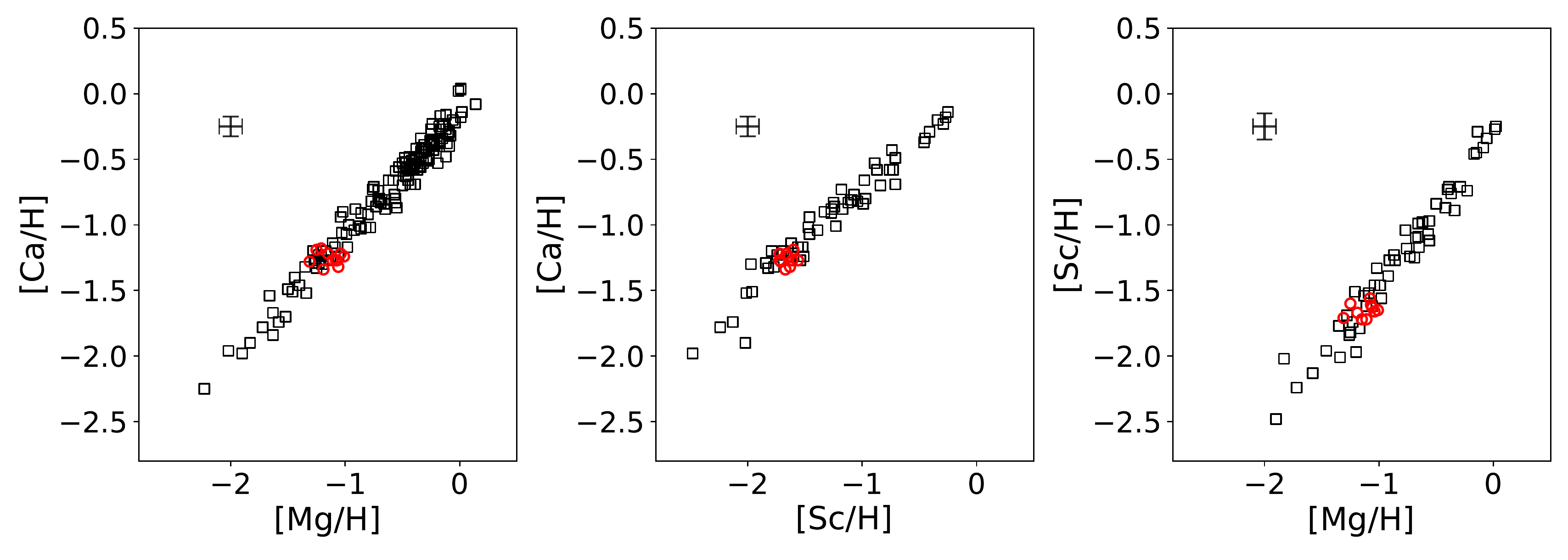}
    \caption{Mg, Ca, and Sc distribution. Black squares and red circles represent field stars from \citet{Gratton2003} and our UVES sample.}
    \label{fig:Gratton2003}
\end{figure*}

We check the same elements as \citet[][]{Carretta2021} using our UVES sample. The Fig. \ref{fig:Gratton2003} shows the [Mg/H], [Ca/H], and [Sc/H] measured in our sample along with fields stars from \citet[][]{Gratton2003}. Mg, Ca, and Sc follow quite clearly the field stars' distribution of similar metallicity, which is consistent with the result found by \citet[][]{Carretta2021} for NGC~6752. We find no evidence of a significant spread in Ca and Sc in this cluster, meaning our results do not support the contribution of explosive polluters to cluster pollution. 

\subsection{Copper}

According to the literature \citep[e.g.,][and references within]{Ernandes2020}, Cu is an iron-peak element mainly produced from a secondary weak s-process in massive stars with a small contribution from other sources such as AGB stars and SNIa.

We studied the Cu line at 5105\AA. While we could only set upper limits in unevolved stars, we derived actual measurements in evolved ones. Among the latter, data show some spread among stars in the same evolutionary stage, which is slightly larger than the error, but there is no trend. According to the literature, copper is under-abundant for stars with metallicity lower than -0.90 dex. Our results show good agreement with field stars at this metallicity.

\begin{figure}
     \centering
     \begin{subfigure}[b]{0.48\textwidth}
         \centering
         \includegraphics[width=\textwidth]{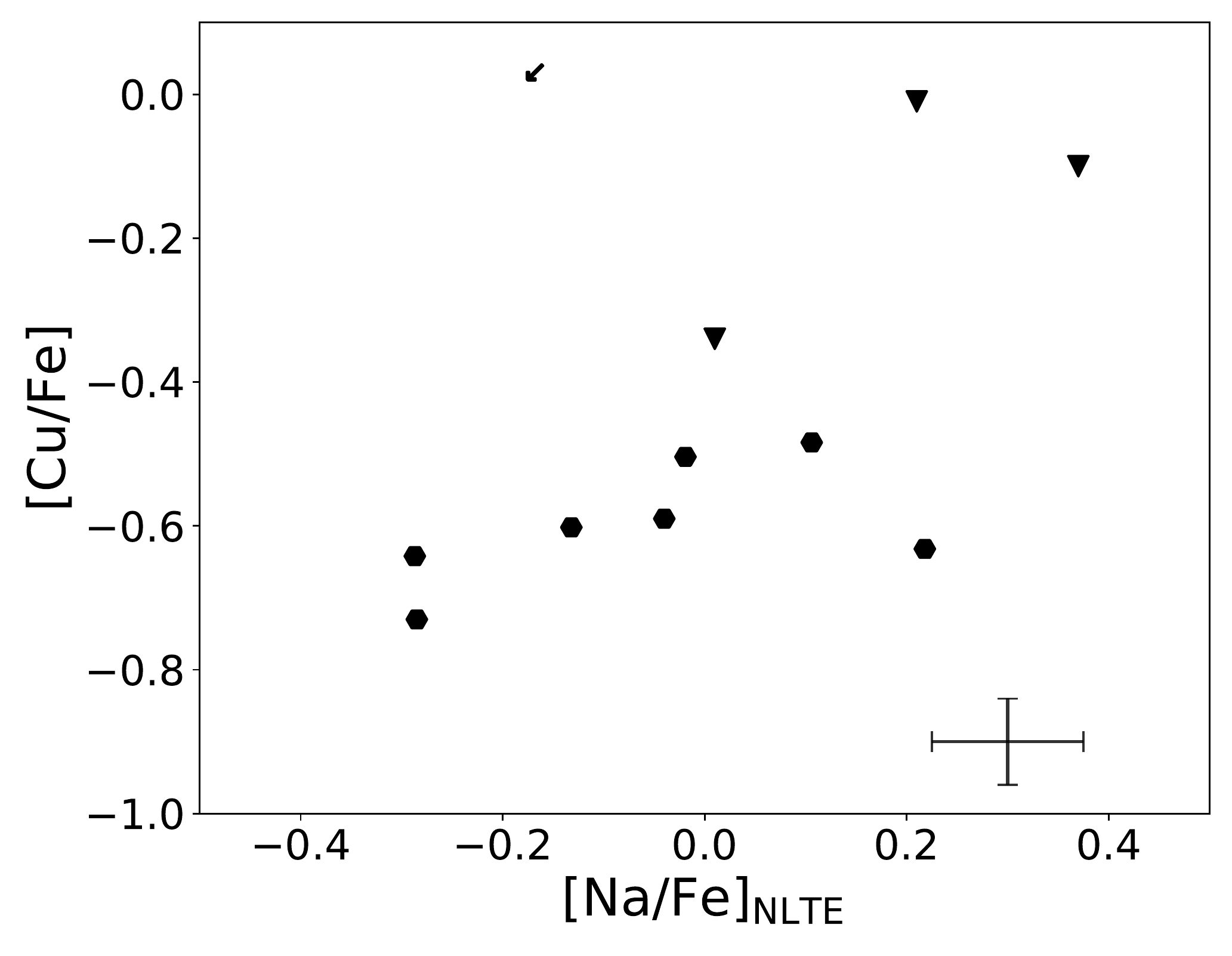}
     \end{subfigure}
     \vfill
     \begin{subfigure}[b]{0.48\textwidth}
         \centering
         \includegraphics[width=\textwidth]{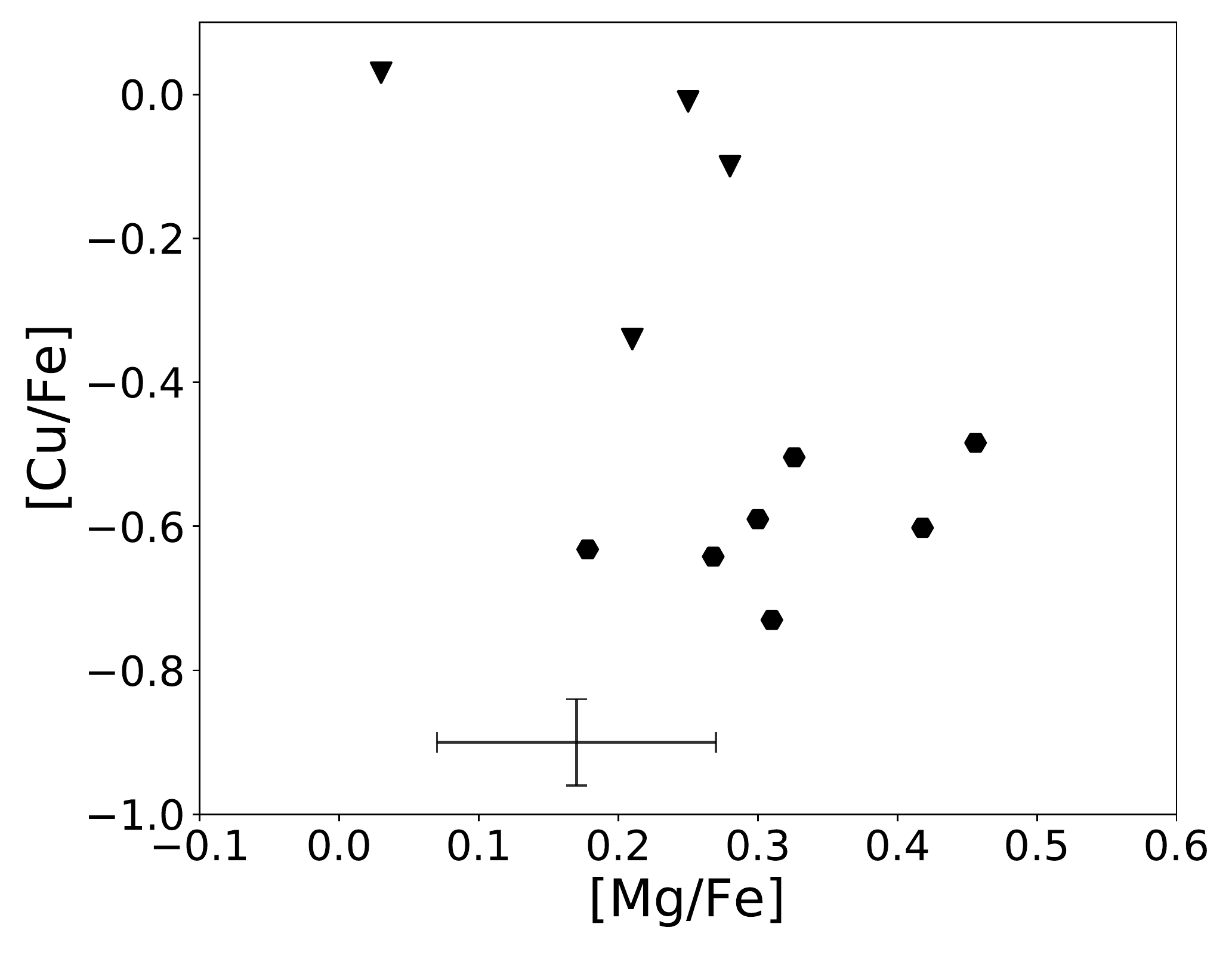}
     \end{subfigure}
        \caption{Relation between [Cu/Fe] and [Na/Fe]$_{NLTE}$ (upper panel), and [Cu/Fe] and [Mg/Fe] (lower panel). Symbols follow the same description as Figure\ref{fig:MgNa}}
        \label{fig:Cu_Na-Mg}
\end{figure}

Figure \ref{fig:Cu_Na-Mg} shows the [Cu/Fe] relation with both [Na/Fe]$_{NLTE}$ (upper panel), and [Mg/Fe] (lower panel). Although the Cu spread is small and within the errors, copper seems to slightly correlate with Na, however, according to a Spearman correlation analysis, the correlation is not significant. In the second, [Mg/Fe] does not display a clear correlation with Cu. Unfortunately, our upper limits are not significant and they do not provide further information. 

\subsection{Yttrium}

Y together with Sr and Zr are known as light-s process elements. They constitute a group of species around the neutron magic number N=50 where arise the first peak of heavy elements \citep{Busso2022}. Unlike Ba, Y production in AGB stars does not strongly depend on masses at this metallicity \citep[2 - 6\msun range;][]{Straniero2014}. The yttrium abundance in field halo stars is found to be slightly sub-solar although with a large scatter \citep{Venn2004,Yong2005}.

\begin{figure}
     \centering
     \begin{subfigure}[b]{\columnwidth}
         \centering
         \includegraphics[width=\textwidth]{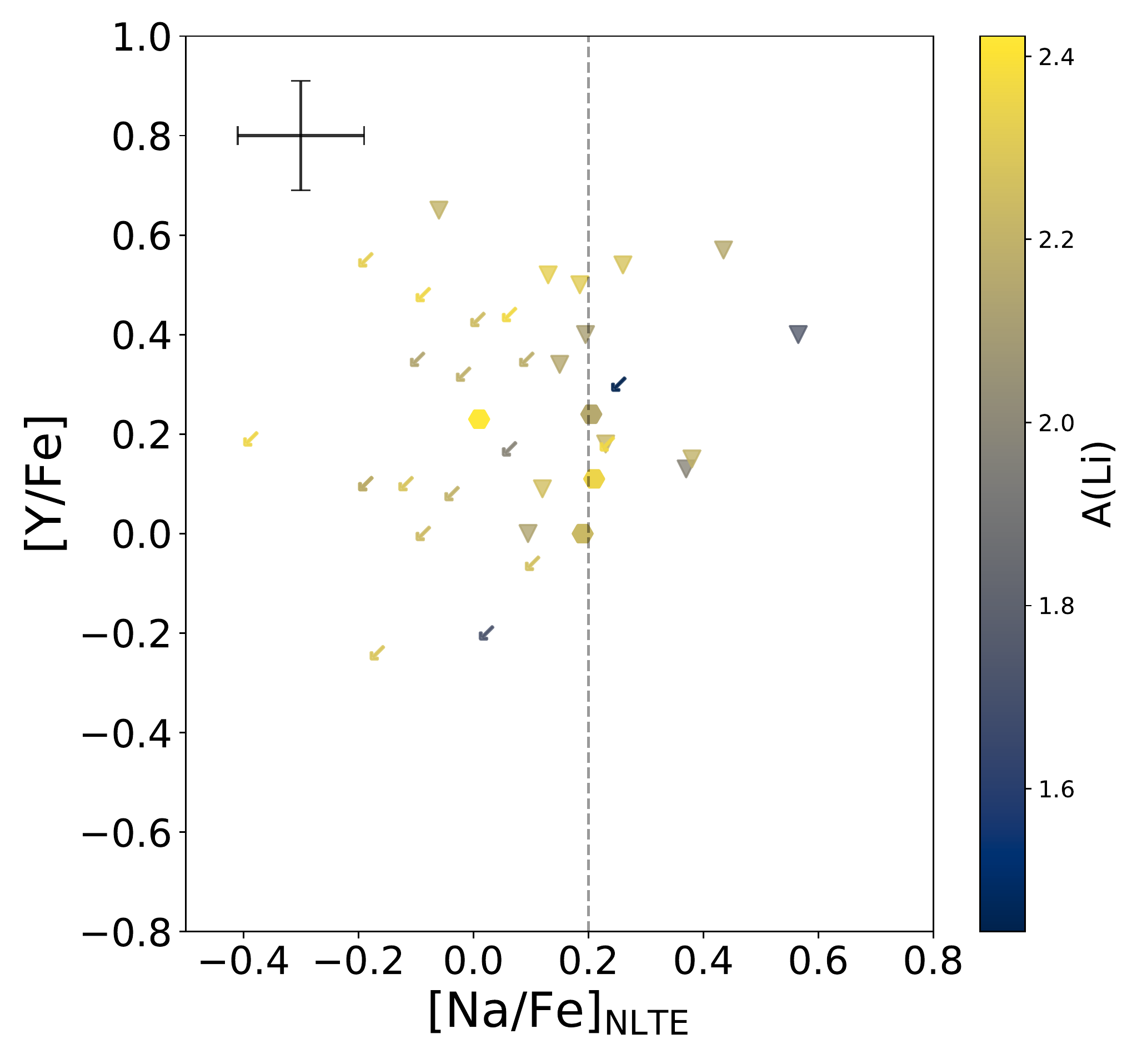}
     \end{subfigure}
     \vfill
     \begin{subfigure}[b]{\columnwidth}
         \centering
         \includegraphics[width=\textwidth]{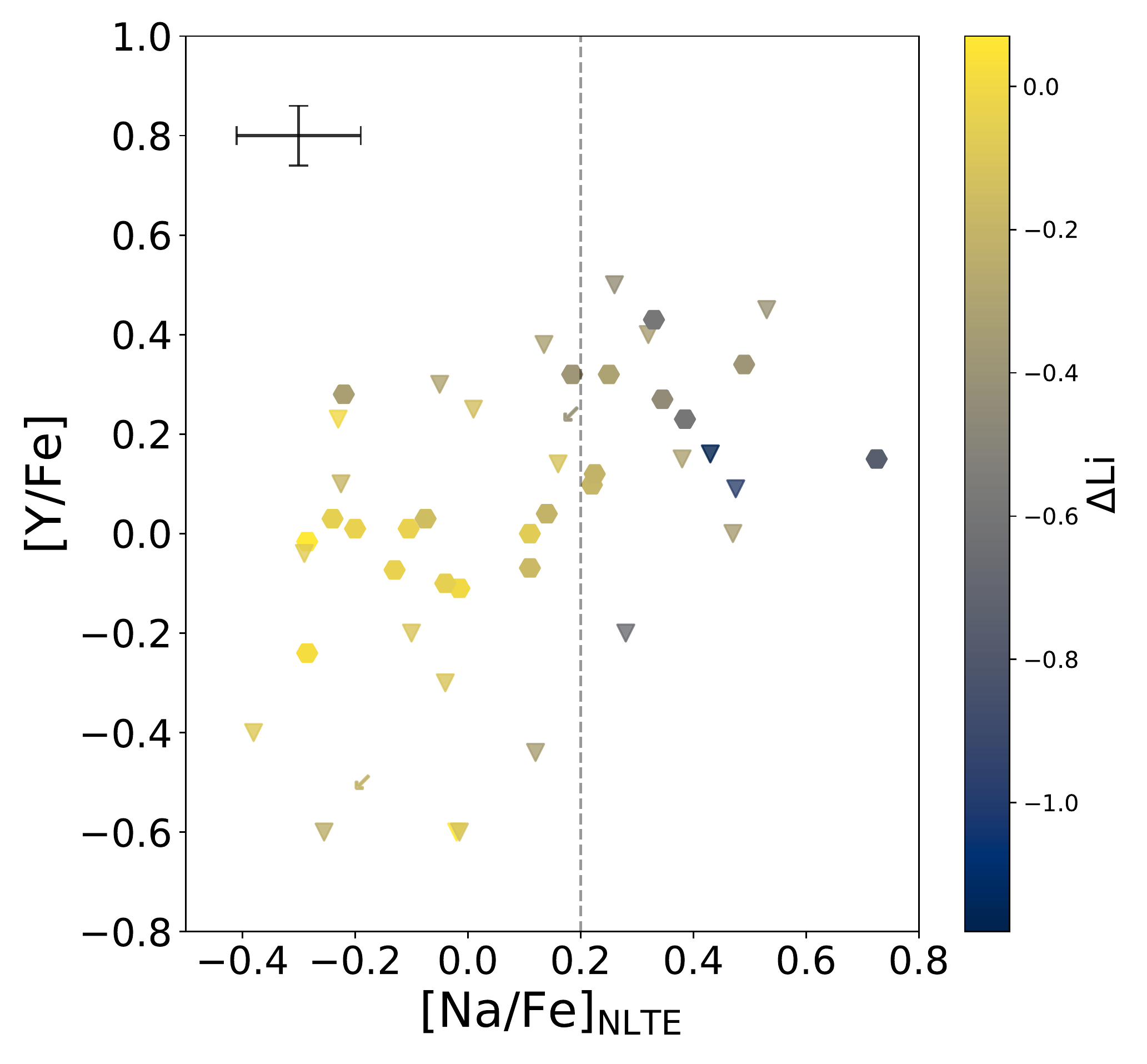}
     \end{subfigure}
        \caption{Weighted Y II abundance using the lines at 5087\AA, 5200\AA, and 5509\AA ~as a function of [Na/Fe]$_{NLTE}$. Results for unevolved, and evolved stars are shown in the upper and lower panels, respectively. Symbols are colour-coded by the star's lithium abundance. Circles and triangles indicate actual measurement and upper limits, respectively. Diagonal arrows indicated the upper limit in both elements.}
        \label{fig:YNa}
\end{figure}

Figure \ref{fig:YNa} shows the distribution of the [Y/Fe]II along with [Na/Fe]$_{NLTE}$ among both unevolved (upper panel), and evolved stars (lower panel). The symbols are colour-coded by their Li/$\Delta$A(Li) abundances. Our results among unevolved stars are dominated by Y II upper limits, making it impossible to probe any trends. The lower panel suggests a positive correlation, with evolved stars belonging to the SG appearing more Y II rich than their FG counterparts. The Spearman correlation ($\sim$0.65),  however, is of poor statistical significance. On the other hand, as colours reflect the Li abundance, a prevalence of blueish (yellowish) colours among Na-rich (Na-poor) stars reflects the finding discussed in Sect \ref{subsec:ONaMgLi}, that Na-rich stars are on average more Li-poor. 

A similar behaviour can be found between Li and Y II among evolved stars. To investigate the reliability of this trend, we selected the five most Li-poor and the five most Li-rich stars among the SG population, to study if they have different production of Y II. We found a constant Y II abundance within the errors among these two groups of stars.

\begin{figure}
	\includegraphics[width=\columnwidth]{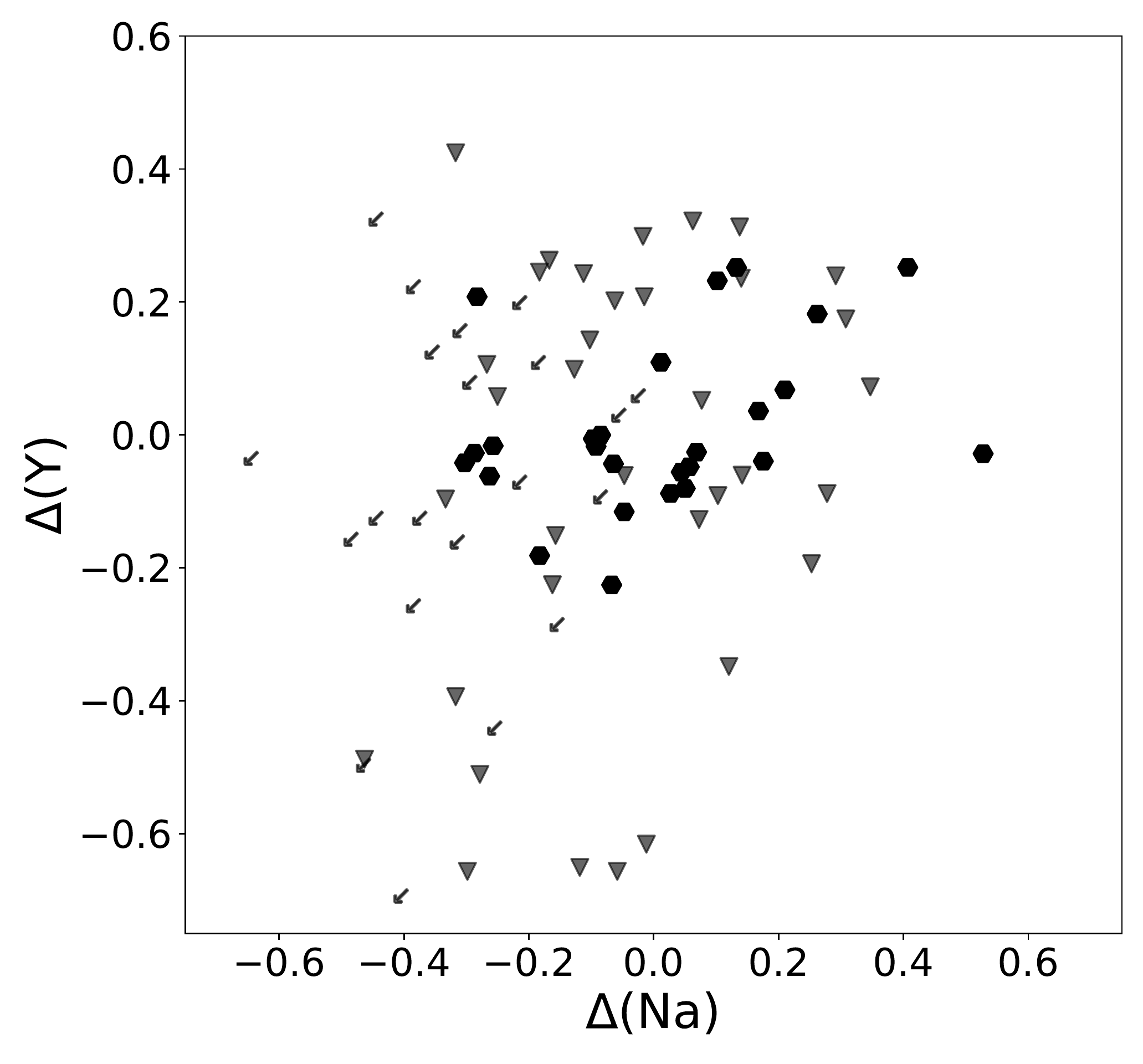}
    \caption{Distribution of $\Delta$[Y/Fe] II and $\Delta$[Na/Fe] in our sample. Symbols follow the description as Fig.\ref{fig:MgNa}}
    \label{fig:Delta_NaY}
\end{figure}

To further analyse the Y II-Na trends among evolved stars, we divided our measurements into four bins. Those bins are given by their Na content, being [Na/Fe]$_{NLTE}$<-0.1 dex, -0.1 dex <[Na/Fe]$_{NLTE}$<0.2 dex, 0.2<[Na/Fe]$_{NLTE}$<0.5 dex, [Na/Fe]$_{NLTE}$>0.5 dex. The corresponding average [Y/Fe]II abundances are 0.00$\pm$0.15 dex, 0.04$\pm$0.15 dex, 0.24$\pm$0.11 dex, and 0.15 dex. They do not show a difference within the errors, therefore it does support a conclusion of correlation.

For sanity check, and to rule out any dependency of Y II and Na on v$_{m}$ (which as discussed in Sect.~\ref{Sec:Data Analysis} might be an issue, at least for Y), we defined $\Delta$Y II ($\Delta$Na) as the difference of the reported [Y/Fe] II ([Na/Fe]) abundance and a linear fit between the [Y/Fe] II ([Na/Fe]) and v$_{m}$. Fig. \ref{fig:Delta_NaY} shows the distribution of these two variables in our sample. A Spearman correlation indicates a weak correlation with low significance. The lack of correlation between Na and Y II suggests that the source responsible for the Na enrichment has quite limited Y II production, if any. 

\begin{figure}
	\includegraphics[width=\columnwidth]{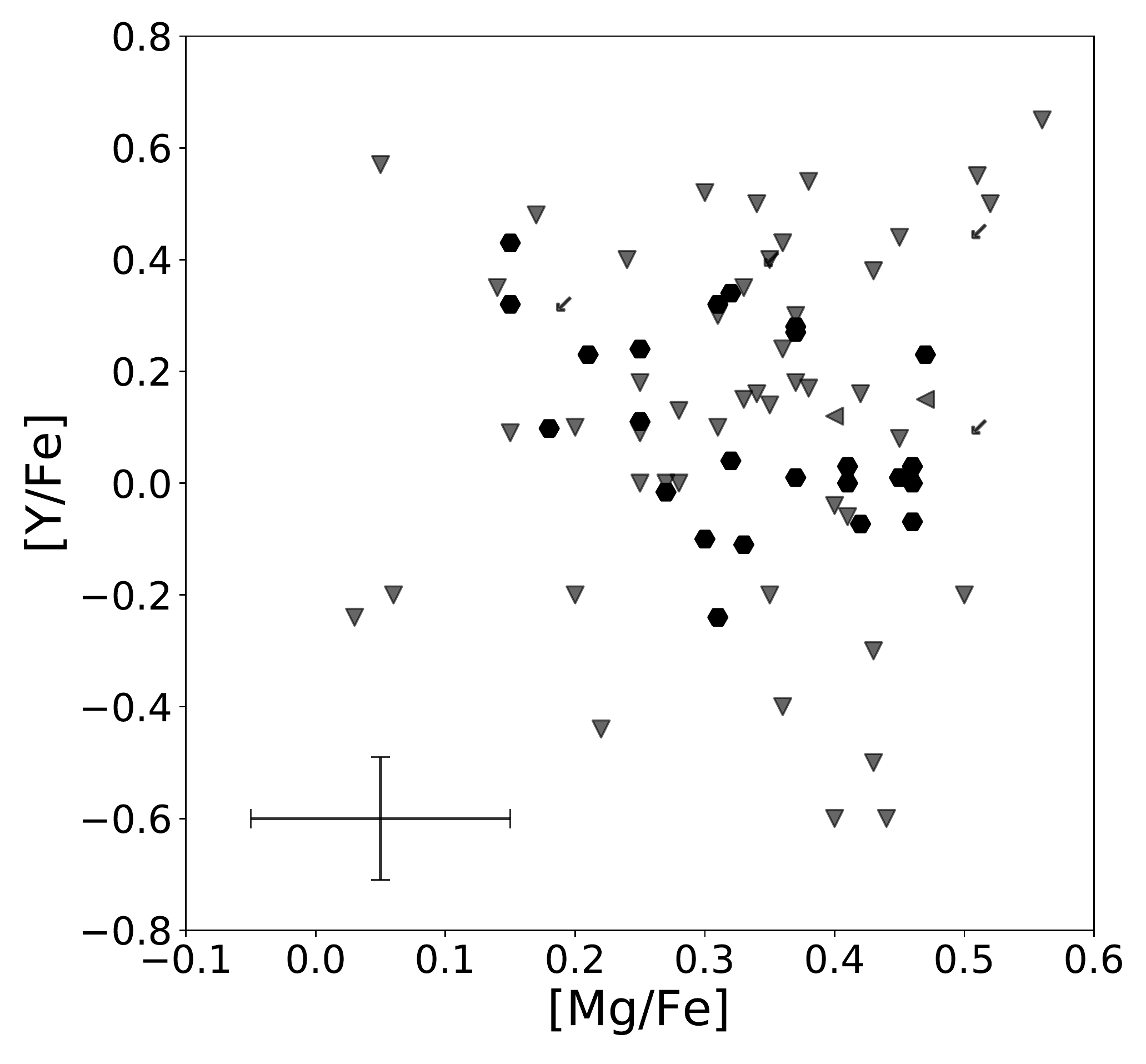}
    \caption{Y and Mg abundance in our sample. Symbols follow the description as Fig.\ref{fig:MgNa}}
    \label{fig:YMg}
\end{figure}

\begin{figure}
	\includegraphics[width=\columnwidth]{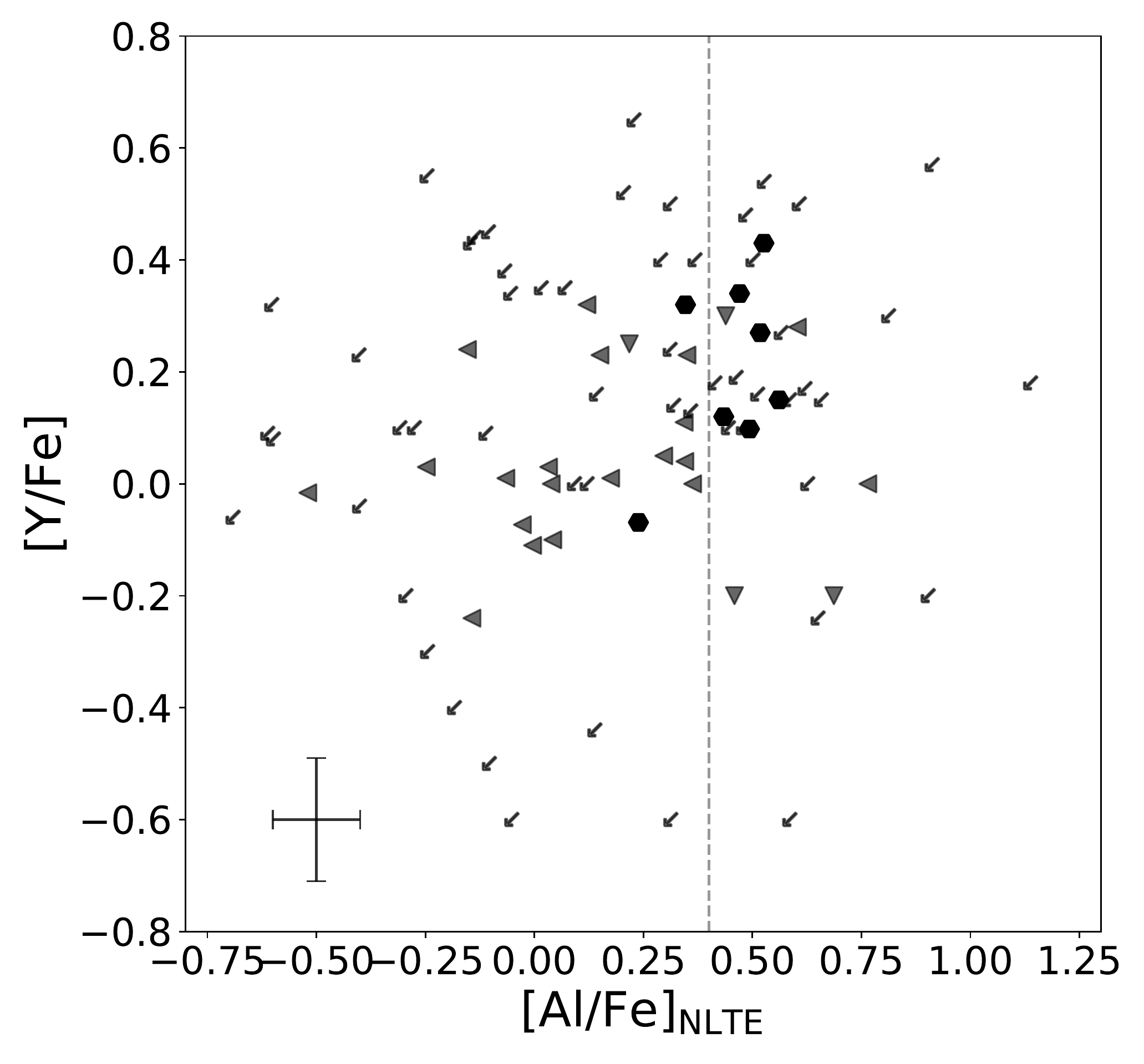}
    \caption{Y abundances as a function of Al. Symbols follow the description as Fig.\ref{fig:MgNa}}
    \label{fig:YAl}
\end{figure}

Fig.\ref{fig:YMg} shows the distribution of [Y/Fe]II along with the Mg abundance. These two species do not display any correlation which agrees with the previous figure. A similar test can be done taking into account the Al abundance (see Fig. \ref{fig:YAl}). Although the figure is dominated by upper limits in both elements, there is once again no clear evidence [Y/Fe]II enrichment in SG with respect to the FG population.

\begin{figure}
     \centering
     \begin{subfigure}[b]{0.49\textwidth}
         \centering
         \includegraphics[width=\textwidth]{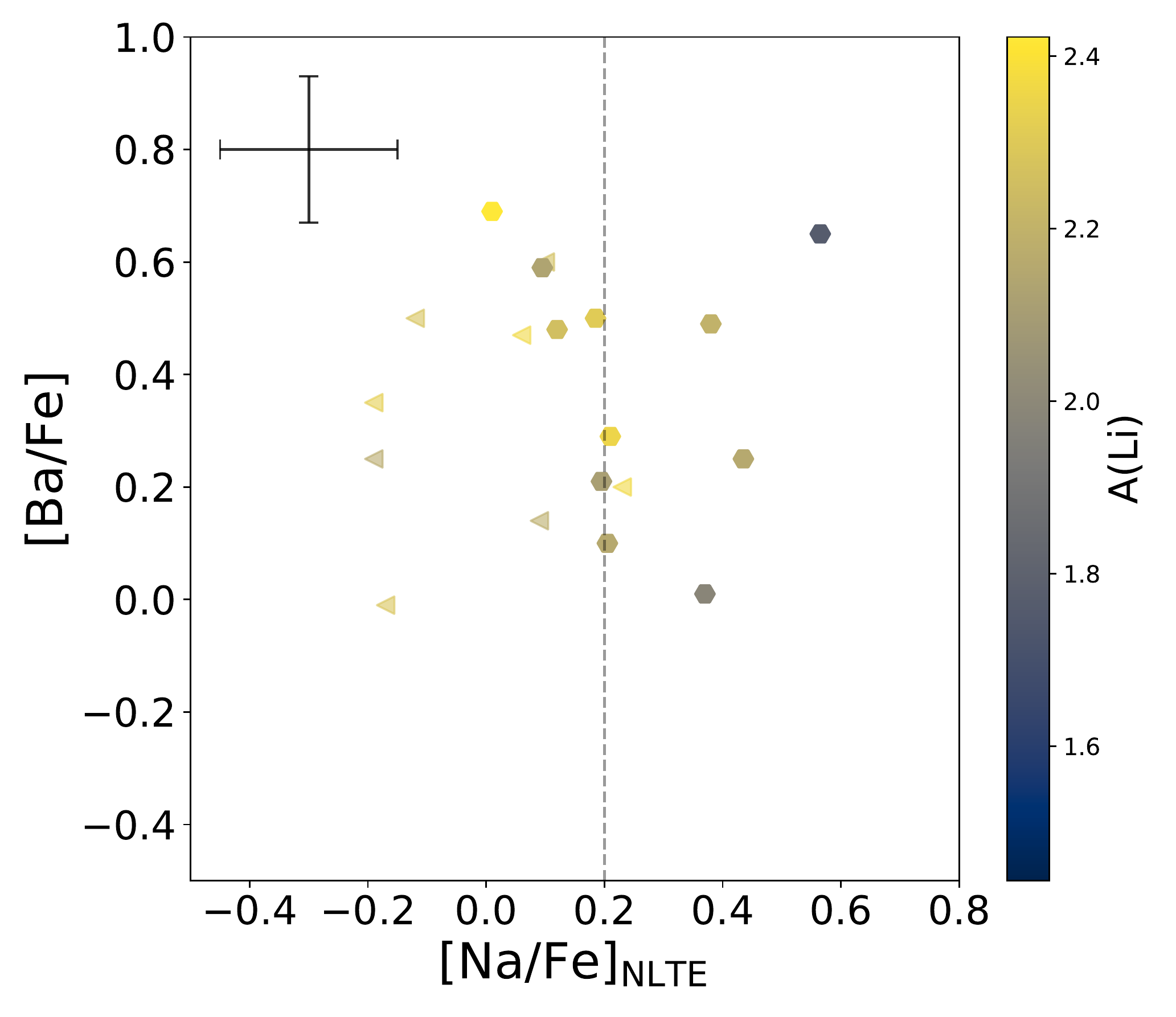}
     \end{subfigure}
     \vfill
     \begin{subfigure}[b]{0.49\textwidth}
         \centering
         \includegraphics[width=\textwidth]{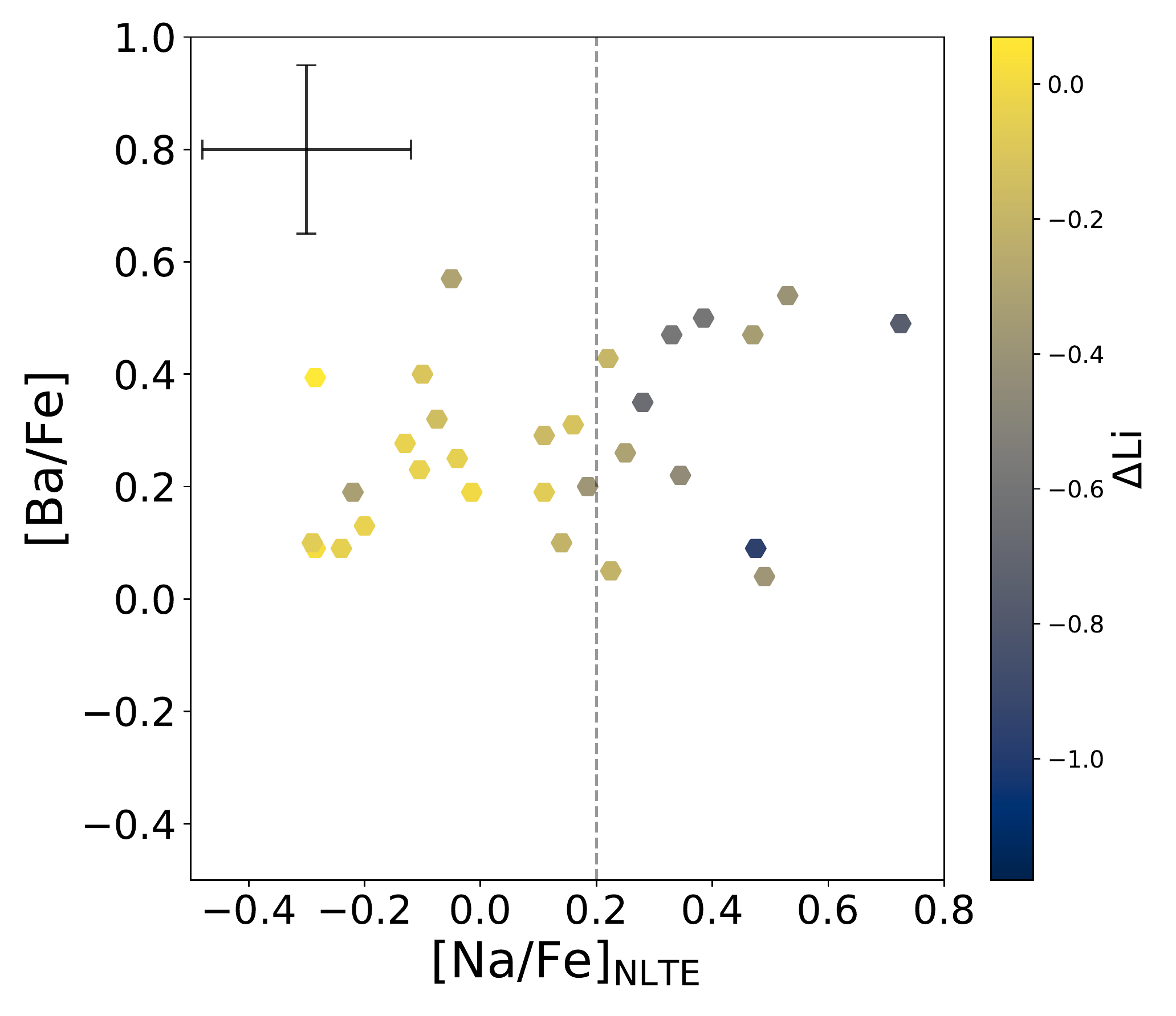}
     \end{subfigure}
        \caption{Weighted Ba abundance using the lines at 5853\AA, 6141\AA, and 6496\AA ~as a function of [Na/Fe]$_{NLTE}$. The panels, symbols and colors follow the same description given in Fig.\ref{fig:YNa}.}
        \label{fig:BaNa}
\end{figure}

\subsection{Barium}

Ba is mainly produced by the main s-process, which happens typically in low-mass ($\sim$1.2 - 4.0\msun ~exact range depending on metallicity) AGB stars during their thermal pulses \citep[][]{Cristallo2015}.

Figure \ref{fig:BaNa} displays the Ba abundance as a function of Na. The figure's colour and symbols follow the same description as Fig.\ref{fig:YNa}. The upper panels show the results for unevolved stars. In both figures, the Ba spread and the Na abundances do not display any obvious correlation. In other words, there is no evidence of a difference in Ba abundance between Na-rich and Na-poor stars. We note, however, that the FG stars are dominated by Na upper limits. Furthermore, the colours do not reveal any obvious correlation between the [Ba/Fe] II and Li abundances. In the lower panels, we present the results for evolved stars, which seems to indicate a possible correlation. However, the Ba II content in the five more Li-poor SG and the five more Li-rich SG stars, does not reflect any correlation considering the associated errors, meaning that the polluters responsible for the Li pollution does not have relevant production of Ba II. 

\begin{figure}
	\includegraphics[width=\columnwidth]{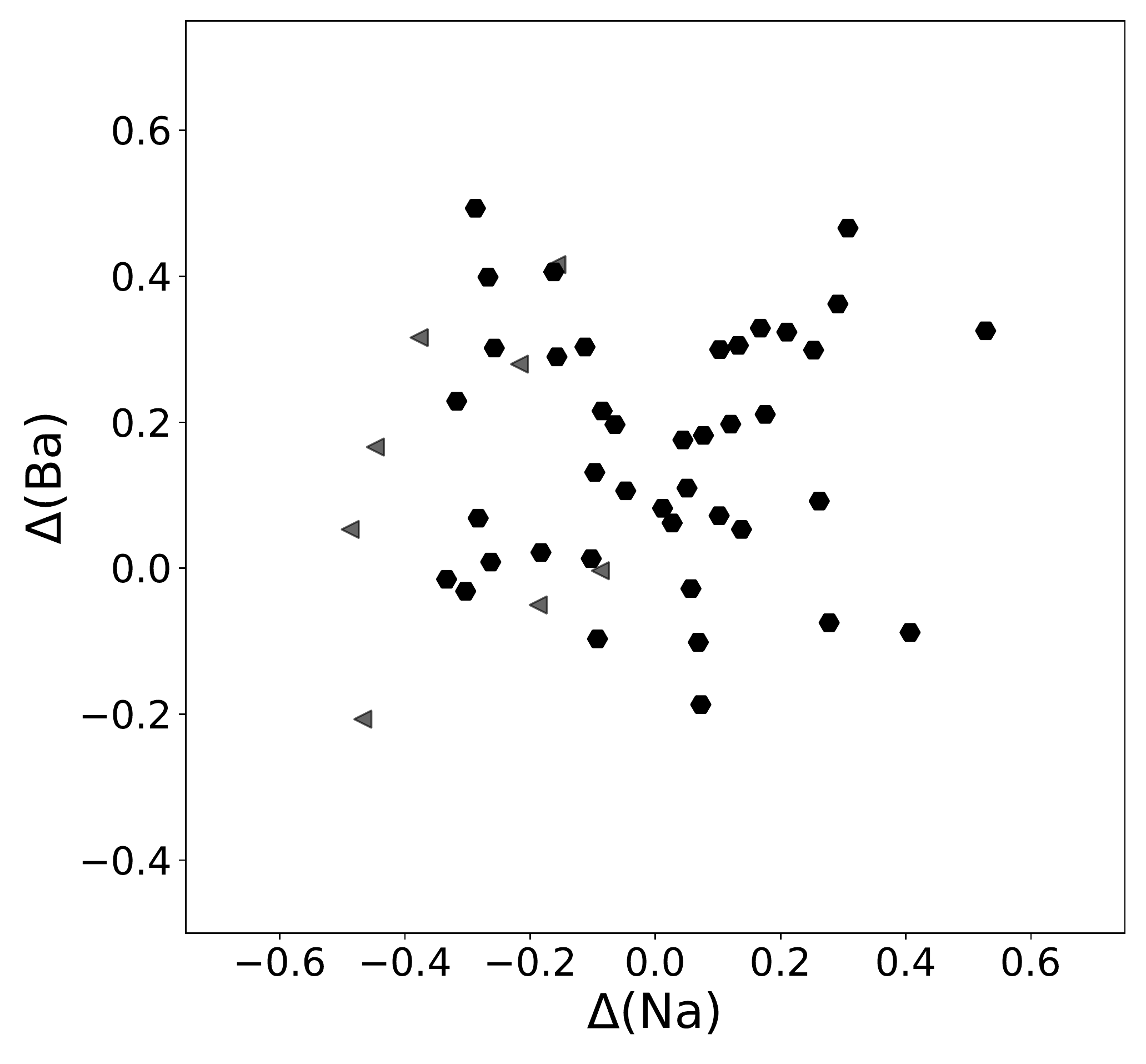}
    \caption{Distribution of $\Delta$Ba and $\Delta$Na in our sample. Symbols follow the description as Fig.\ref{fig:MgNa}}
    \label{fig:Delta_NaBa}
\end{figure}

Furthermore, as was done previously with Y II, we compared the mean Ba abundance among both unevolved and evolved FG and SG stars. No systematic trends in Ba with any of these groups are seen. Also, we defined $\Delta$[Ba/Fe] II as was done $\Delta$[Na/Fe], and we compared them in Fig. \ref{fig:Delta_NaBa}. The results confirm that there is no trend with Na, suggesting that the possible trend observed in Fig. \ref{fig:BaNa} might arise from dependencies from microturbulence and that the first and second generation of stars displays a similar Ba abundance and the Na producer is not related to the production of Ba.

\begin{figure}
	\includegraphics[width=\columnwidth]{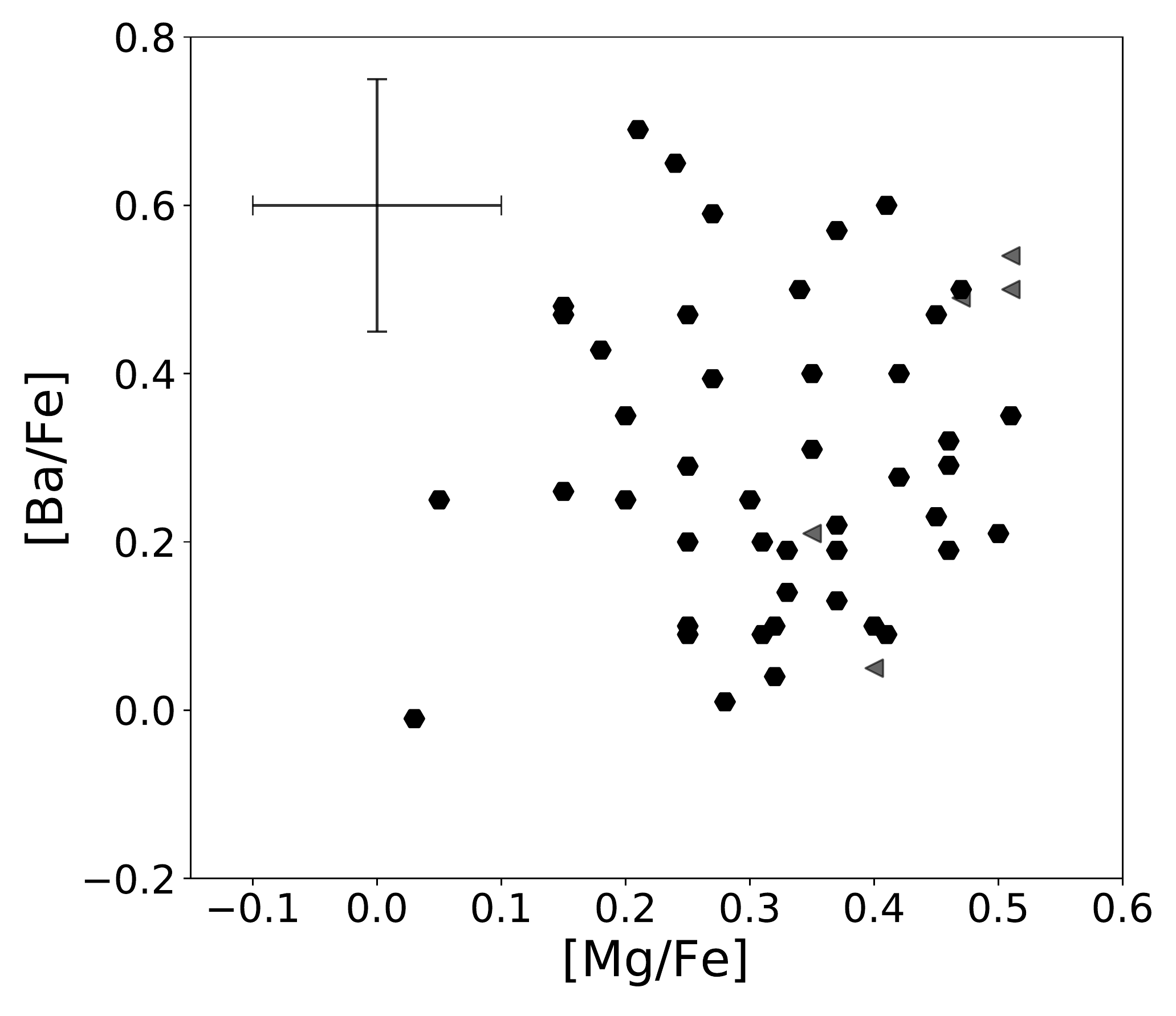}
    \caption{Ba abundances along with the Mg content in our sample. Symbols follow the description as Fig.\ref{fig:MgNa}}
    \label{fig:BaMg}
\end{figure}

\begin{figure}
	\includegraphics[width=\columnwidth]{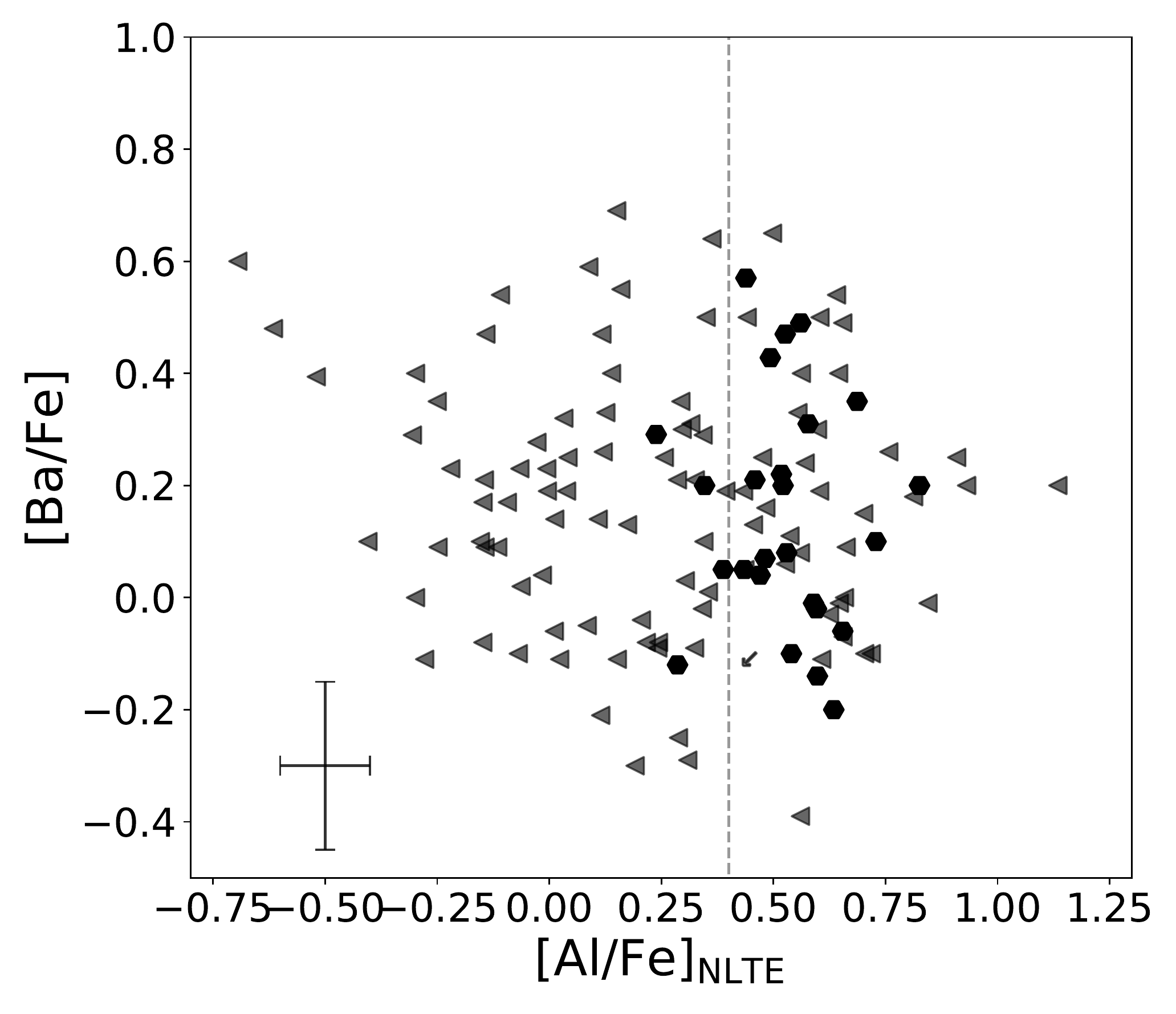}
    \caption{Ba abundances as a function of Al in NGC~6752. Symbols follow the description as Fig.\ref{fig:MgNa}}
    \label{fig:BaAl}
\end{figure}

Figure \ref{fig:BaMg} displays the Ba abundance of our sample as a function of the Mg content. It is clear there is no correlation between these two elements, then there is no evidence indicating they were synthesised at the same site. Fig. \ref{fig:BaAl} shows the distribution of Ba along with the Al abundance. The figure is dominated by Al upper limits, however, the reported measurements do not reveal any correlation, then one generation does not seem to be Ba enriched with respect to the other. 

On the other hand, we found evidence of a Ba spread in both unevolved and evolved stars, which given the reported error is mildly significant. The spread seems to be constant in the different stellar populations. The spread in heavy elements has been reported in a few GCs before, e.g., NGC~7089, and M~22. 

\begin{figure*}
     \centering
     \begin{subfigure}[b]{0.335\textwidth}
         \centering
         \includegraphics[width=\columnwidth]{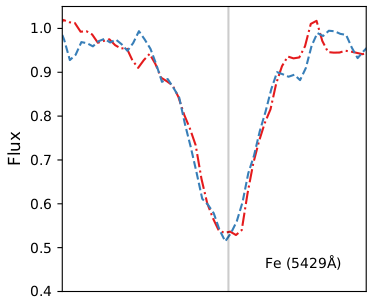}
     \end{subfigure}
     \hfill
     \begin{subfigure}[b]{0.32\textwidth}
         \centering
         \includegraphics[width=\columnwidth]{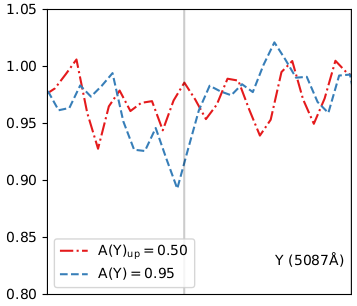}
     \end{subfigure}
     \hfill
     \begin{subfigure}[b]{0.31\textwidth}
         \centering
         \includegraphics[width=\columnwidth]{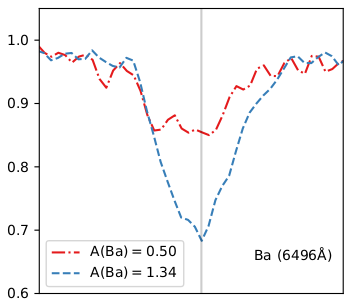}
     \end{subfigure}
        \caption{Line-by-line comparison between two members with similar stellar parameters ($\Delta$T$_{\mathrm{eff}}\sim$80 K). The referred lines are indicated in the right upper corner of each panel. The gray solid line indicates the position of the line centre.}
        \label{fig:Spec_Comp1}
\end{figure*}

We noted a star with particularly high Ba, and we performed a direct comparison to another star with similar stellar parameters, but different Ba content. Figure \ref{fig:Spec_Comp1} shows a line-to-line comparison between these two stars for Y and Ba at 5087\AA and 6496\AA respectively. As can be seen from the figure, the Ba lines are quite different, consistently with the A(Ba) abundances that have been measured in the two stars, arguing in favour of a real difference in Ba rather than uncertainties in the parameters. The abundance of Y II seems to follow the same pattern as Ba II, although with a smaller effect. As we could only place upper limits for Cu and Eu, no meaningful comparison can be performed for these species. 

The measured abundances of s-process elements and the low Eu upper limit, suggest an enrichment by the s-process. This anomalous n-capture abundance could be explained by it being a CH-stars. Those stars, and their metal-rich counterparts Ba-stars, have been the object of a number of literature studies. In GCs, Ba-rich stars are mostly FG \citep[e.g.,][]{dorazi_gratton2010} and they are thought to be part of binary systems, which is consistent with the fact that binary fraction is higher in FG than in SG stars \citep[e.g.,][]{Lucatello2015}. In these binary systems, the primary was a star with 1.5-4.0\msun ~which evolved long ago and is now a faint white dwarf. Such a star, after its AGB phase, transferred mass to the secondary -- the star we are now observing --, enriching the atmosphere of the latter with products of the AGB shell nucleosynthesis. The result is that the surface composition is enriched in s-process elements, but also C.

Our Ba-enhanced star has a high A(Ba)=0.92 dex ([Ba/Fe]=0.52 dex), and the expectation is that it would also be C-rich. Given the temperature/metallicity of the object, the CH G-band at 4300\,\AA~ is the best feature to check for C enhancement. Our spectral coverage, however, does not allow that. The C$_2$ Swan band at $\sim$5250\,\AA~ requires a quite high C enhancement to be detectable, and a visual inspection does not reveal a clear presence of this band in the spectrum of the star. Given the atmospheric parameters of the stars and the spectral SNR in the region, we estimated in [C/Fe]=$1.7$\,dex the minimum C-abundance to lead to a detectable C$_2$ Swan band. Then, we compared these abundances with the expected AGB yields for a star with 1.50\msun from F.R.U.I.T.Y models. 
The predicted [Ba/Fe] and [C/Fe] abundances should be about 1.45 dex, and 2.30 dex, respectively. That means the [Ba/Fe] coming from the material of the companion of our CH-star should have been diluted by about 0.9 dex to get to the observed [Ba/Fe] abundance of 0.52\,dex. If we consider the same dilution for C, the [C/Fe] expected in our star should be around 1.4 dex, which would result in an undetectable Swan band in our spectra. Therefore, the lack of Swan band is not an argument against the target being a CH-star, which remains the most likely cause of the observed s-process enhancement.

We attempted to measure Eu abundance from the 6645\,\AA line, however, the feature was too weak at the atmospheric parameter of the sample and SNR of the observed spectra. We hence only derived upper limits of scarce significance.

\subsection{n-capture elements distribution}

\begin{figure}
	\includegraphics[width=\columnwidth]{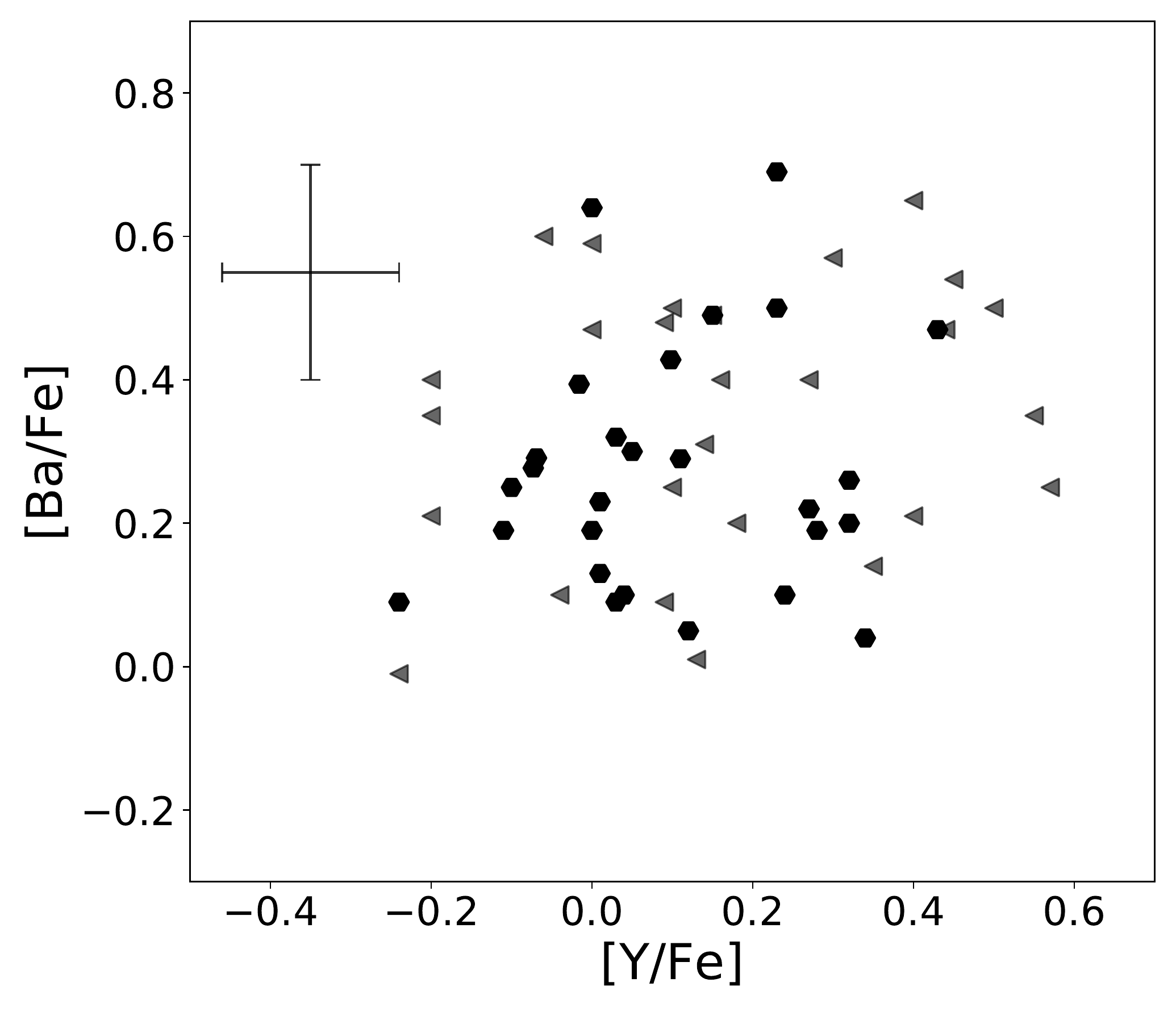}
    \caption{Barium abundances as a function of [Y/Fe] II. Symbols follow the same description as the previous figures.}
    \label{fig:BaY}
\end{figure}

Figure \ref{fig:BaY} shows the [Y/Fe]II abundance versus Ba. The distribution presents the dispersion discussed previously, but there is no evidence of any correlation. It is worth noting that, although AGB stars are the main ones responsible for the s-process nucleosynthesis, Y and Ba can be produced by stars of different stellar masses, then a correlation between these two species is not granted when a range of AGB masses is involved in the pollution.

\subsection{n-capture elements and AGB predictions}

In the scenario where AGB stars are the polluters responsible for the abundance variations, we can take advantage of the measured Y  and Ba abundances to explore the mass of the AGB involved in the process. Under the simplistic approximation that polluters are all identical, that the composition of the diluting gas is the same as the FG stars  and that the Na and Y content in FG stars is negligible with respect to the Na and Y content coming from the AGB ejecta, we expect that Na and Y would vary proportionally to the yields of the polluter.
We robustly estimate the overall increase in Na, and we calculated the difference between averaging the five Na-richest and five poorest measured abundances. An analogous procedure was applied to Y, deriving 1.05$\pm$0.13 dex, and 0.88$\pm$0.13 dex for [Na/Fe] and [Y/Fe] II, respectively.

These quantities were compared to results from nucleosynthetic models from the FUll-Network Repository of Updated Isotopic Tables \& Yields\footnote{\url{http://193.204.1.214/modelli.pl}} \citep[F.R.U.I.T.Y;][]{Cristallo2011}, providing theoretical predictions for the yields of a range of AGB masses, from 1.5 to 6\msun. Table \ref{tab:models} displays the final composition of diverse elements for a given AGB mass. The model is for ~Z=0.0003, [$\alpha$/Fe]=0.5 dex, and a standard C-pocket. The 4\msun model actually predicts that the Y production is actually slightly higher than that of Na, something that is not consistent with our data, where the estimated Na increase is equivalent to Y, considering the errors. The 5 and 6\msun~ model predictions are on the other hand fully consistent within the errors in our findings. 

We cannot extend the exercise to higher masses for the lack of available models at the appropriate mass-metallicity combination. Some qualitative insight could potentially be gained by comparing the FRUITY models with others for larger masses, like e.g. \citet{Karakas2018} which extends to 7\msun but only at higher metallicity. However, their predictions for Na, Y and Ba abundances at a similar metallicity ([Fe/H]$=-0.7$) and in the range of overlapping masses (5 and 6\msun) are only in moderately good agreement, suggesting extreme caution in even qualitative speculations on the behaviour of the predictions for Na, Y and Ba at higher masses.

\begin{table}
\centering
\caption{The final composition of the AGB star for each species from F.R.U.I.T.Y models for different stellar masses.} 
\label{tab:models}
\begin{tabular}{cccc}
\hline
\hline
Mass & {[}Na/Fe{]} & {[}Y/Fe{]} & {[}Ba/Fe{]}\\
\hline
4M$_{\odot}$ & 0.81    & 0.89  &   0.99   \\
5M$_{\odot}$ & 0.93    & 0.94  &   0.77   \\
6M$_{\odot}$ & 0.98    & 0.73  &   0.50   \\
\hline
\hline
\end{tabular}
\end{table}

Similar reasoning can be applied to Na and Ba with the estimated overall [Ba/Fe] II increase of 1.01$\pm$0.18 dex. We found a similar Ba and Na production in our sample, however, none of these models predict such a pattern. Models of 5\msun and 6\msun show a comparable Na increase to our result, but a much lower Ba increase. Our findings can not be attributed to a single AGB polluter of a single mass.

It is interesting to compare this result with Li, as its content in FG stars is only expected to change because of the evolutionary stage of the star. Li abundance determinations should not be affected by small Fe variations, hence it should be quite robust with respect to a potential Fe spread. The 4 stars previously mentioned are evolved stars, so we compared their $\Delta$Li abundances as defined by \citetalias{Schiappacasse2021}, which ranges from -0.17$\pm$0.05 to 0.07$\pm$0.05. This Li spread should be further investigated in a larger sample, but it could suggest a spread in elements other than Fe in FG stars, which could also be a potential explanation for part of the spread in the SG ones. Similar Li spread in FG stars have been found previously in other GCs: \cite{dorazi2014} reported a Li difference of $\sim$0.25 dex among their FG sample in NGC~6218 and NGC~5904, and later in the GCs NGC~2808 and NGC~362 \citep{Dorazi2015}.

\section{Summary and Conclusion}
\label{Sec:Conclusion}

We studied a sample of 158 stars in the GC NGC~6752, from the TO to the RGB bump. In particular, we analysed transitions for Na, Mg, Ca, Sc, Cu, Y, and Ba deriving abundances or meaningful upper limits whenever possible. Furthermore, we attempted to measure the abundances for O and Eu, deriving only upper limits for the whole sample.
We extended the analysis done by \citetalias{Schiappacasse2021} studying the potential relation between our abundances of the n-capture elements and their results on Li and Al. We aimed to derive constraints on the nature of the polluter(s) responsible for the chemical features found in SG stars in NGC~6752. 

The abundances of elements such as Mg, Ca, and Sc showed that our sample follows the expected field distribution at the metallicity of the cluster. Ca and Sc show no significant spread in the sample confirming previous results in the literature. As it is expected, Mg measurements display a larger spread, which anti-correlates with Na and Al results, which are in good agreement with the literature. This behaviour is a typical signature of the MSP phenomenon produced by polluters undergoing nucleosynthesis typical of high temperatures of H-burning. Still, the temperature did not reach the point to involve heavier elements such as Ca or Sc. Unfortunately, because we could only set upper limits in O, we could not probe the well-studied Na-O anti-correlation. 

Cu shows a quite constant abundance within the error and follows the field pattern at the same metallicity. Cu abundances do not display any correlation either with Na or Mg or with heavier elements.

As a representative of the first peak s-process element, [Y/Fe]II was analysed and we detected a strong correlation with the adopted v$_{m}$, which we minimised by using the mean weighted by their errors of all 3 Y II lines detectable in our spectra. We found a mildly significant [Y/Fe]II spread in our sample considering the associated errors. The analysis of Y II with Li, Na, Mg, and Al does not reveal any correlation, meaning that the results are not compatible with considerable pollution of Y II in SG stars.

As a representative of the second peak s-process elements, we measured the Ba content in our sample, which shows a marginally significant spread in its distribution. The study of Ba together with Na, Mg, and Al do not show any difference between FG and SG stars, meaning that considerable Ba production in the polluters can be excluded. Additionally, we reported a FG star with particularly high Ba and Y abundances, likely a CH star.

We explored the possible iron spread in our FG sample, as suggested by \citet[][]{Legnardi2022}.
Although, we found a similar Fe spread among the FG stars of our sample. Differential analysis of four FG stars with similar stellar parameters ($\Delta$T$_{\mathrm{eff}}$=25 K) showed a Fe spread ($\Delta$[Fe/H]=0.07 dex), which is within the associated errors. Interestingly, those stars showed a significant Li spread, which should have been present since the cluster's birth. Nevertheless, this hint should be taken with caution due to our small sample. 

The presence of Li-rich SG stars in our sample is clear evidence of the intermediate-mass AGB star contribution to cluster pollution. According to models, these stars are expected to have some production of s-process elements, which could produce also the spread that we found in these species. However, the comparison with our results reflects that the pollution of Y II and Ba II should be quite modest if any, and it could possibly be consistent with the predictions from 6\sun~ and 5\msun AGB stars, respectively. However, we cannot exclude the contribution of AGB stars of higher masses (up to 8\msun), due to the lack of models predicting the heavy element production in these stars. Furthermore, we did not find any relation in the Y II and Ba II content between the Li-rich SG and Li-poor SG stars suggesting that the AGB stars minting the Li necessary to reproduce the observed patterns do not yield considerable amounts of Y and Ba.

\section*{Acknowledgements}

We thank our referee, Chris Sneden, for his helpful comments and recommendations.

J.S-U and his work were supported by the National Agency for Research and Development (ANID)/Programa de Becas de Doctorado en el extranjero/DOCTORADO BECASCHILE/2019-72200126.

This work was partially funded by the PRIN INAF 2019 grant ObFu 1.05.01.85.14 ('Building up the halo: chemo-dynamical tagging in the age of large surveys', PI. S. Lucatello)

%%%%%%%%%%%%%%%%%%%%%%%%%%%%%%%%%%%%%%%%%%%%%%%%%%
\section*{Data Availability}

The spectra used in the article are available on the ESO Science Archive Facility\footnote{\url{http://archive.eso.org/wdb/wdb/adp/phase3\_spectral/form}} under the ID programs 077.D.0246(A), 079.D-0645(A), 081.D-0253(A), and 083.B-0083(A).
The data provided by \citet[][]{Gruyters2014} can be found from the same site under the ID programs 079.D-065(A) and 081.D-0253(A). Str\"{o}mgren photometry is from \citet[][]{Grundahl1999} and it was kindly provided by the author. Last, Gaia eDR3 data are available on \url{https://gea.esac.esa.int/archive/}.



\bibliographystyle{mnras}

% Don't change these lines
\bsp	% typesetting comment
\label{lastpage}
\end{document}